\documentclass[tighten,twocolappendix,twocolumn]{aastex7}

\usepackage[T1]{fontenc}
\usepackage{amsmath}
\usepackage{xspace}
\usepackage{enumitem}
\usepackage{soul}
\usepackage{xcolor}

\newcommand{\bhspin}{a_*}
\newcommand{\sgra}{{Sgr A$^*$}\xspace}

\newcommand{\athenak}{{\tt AthenaK}\xspace}

\begin{document}

\shorttitle{The SANE, the MAD, and the Chimera}
\shortauthors{Wong \& Stone}

\title{The SANE, the MAD, and the Chimera}

\author[0000-0001-6952-2147]{George N.~Wong}
\email{gnwong@ias.edu}
\correspondingauthor{George N.~Wong}
\affiliation{School of Natural Sciences, Institute for Advanced Study, 1 Einstein Drive, Princeton, NJ 08540, USA}
\affiliation{Princeton Gravity Initiative, Princeton University, Princeton, New Jersey 08544, USA}

\author[0000-0001-5603-1832]{James~M.~Stone}
\email{jmstone@ias.edu}
\affiliation{School of Natural Sciences, Institute for Advanced Study, 1 Einstein Drive, Princeton, NJ 08540, USA}

\begin{abstract}
Non-radiative black hole accretion flows are commonly classified by their magnetic flux state, with standard and normal evolution (SANE) disks and magnetically arrested disks (MADs) marking the usual weak- and strong-flux regimes. We compare three-dimensional general relativistic magnetohydrodynamics simulations of a weakly magnetized SANE flow, a standard MAD, and a Chimera flow fed by a different reservoir of mass, angular momentum, and coherent magnetic flux. The Chimera reaches a MAD-level horizon magnetic flux and launches a powerful electromagnetic jet during an extended non-eruptive interval, showing that a flow can maintain large horizon flux and jet power without sharing the standard MAD's bursty horizon-flux variability, mass-flow distribution, or inner-flow morphology. In the SANE flow, we show that radial support is primarily hydrodynamic and provided by gas pressure gradients, whereas in MAD flows, magnetic pressure and tension enter the radial force budget at comparable order and help regulate the inner flow dynamics. The Chimera remains distinct from the standard MAD in its density structure, funnel-wall geometry, mass-flow channels, radial force budget, and angular-momentum transport throughout the analyzed evolution. We therefore argue that MAD-like behavior is not captured by any single diagnostic, but by a dynamical coupling among horizon flux, jet power, magnetic support, Maxwell transport, surface-layer flow, disk morphology, and eruption activity. The Chimera shows that these outcomes can be separated by accretion history and magnetic-flux supply.
\end{abstract}


\section{Introduction}

Magnetic fields play a central role in the dynamics of black hole accretion flows. They destabilize differentially rotating gas through the magnetorotational instability, sustain turbulent angular momentum transport, shape the structure of geometrically thick radiatively inefficient accretion flows, and mediate the extraction of energy into winds and relativistic jets. The dynamical importance of the magnetic field is often expressed through the language of standard and normal evolution (SANE) and magnetically arrested disk (MAD) accretion. In SANE flows, the horizon-threading magnetic flux remains modest and accretion proceeds through a turbulent, quasi-steady, matter-dominated disk. In conventional MAD flows, accreted large-scale magnetic flux accumulates until magnetic forces near the black hole become dynamically important, producing strong jets, sub-Keplerian orbital motion, intermittent accretion streams, and magnetic flux eruptions \citep{bisnovatyi_1974_madstar,igumenshchev_2003_mad,narayan_2003_mad,narayan_2012_sane,sadowski_2013_sane}.

Here we focus on geometrically thick, radiatively inefficient disks, where the accretion power is not primarily released as disk radiation and can instead be stored in hot plasma, carried inward, or redirected into winds and jets \citep{rees_1982_Origin_radio_jets,narayan_1995_ADAF,ho_2008_llagn,yuan_2014_review}. In this setting, the magnetic state of the flow concerns both local plasma dynamics and the large-scale coupling between black holes, their fuel supply, and their outflows. The SANE/MAD vocabulary is useful because it gives a compact name to a broad reorganization of the inner accretion flow. In widely used Fishbone--Moncrief \citep[FM;][]{fishbone_1976_torus} torus simulations, many familiar MAD signatures tend to appear together: large horizon-scale magnetic flux, strong electromagnetic jet power, magnetically supported inner disks, faster and more intermittent accretion, nonaxisymmetric low-density cavities, and magnetic flux eruptions \citep[e.g.,][]{tchekhovskoy_2011_mad,wong_2022_patoka,narayan_2022_jetsurvey,dhruv_2025_surveyv3}.

But this vocabulary raises a basic interpretive question: what does it mean for an accretion flow to be magnetically arrested? The answer cannot be reduced to a simple time-averaged accretion rate or to a single effective stress prescription, because the density normalization in ideal GRMHD is arbitrary and in the SANE/MAD baseline, the MAD drains its torus more rapidly \citep{wong_2022_patoka,dhruv_2025_surveyv3,wong_2025_mixing}. Rather, the arrested label is better understood as a statement about magnetic regulation of the inflow. This regulation may occur episodically, when flux eruptions displace gas and temporarily impede accretion, or more continuously, as magnetic pressure, tension, and torques reshape the force balance, angular-momentum transport, and flow geometry.

The route by which a flow reaches strong magnetization can therefore affect which MAD-like signatures appear together. We compare the standard FM SANE and MAD baselines with a third flow initialized from a Chakrabarti torus and threaded by a large-scale vertical field \citep{chakrabarti_1985_thickdisk}. This setup supplies mass, angular momentum, and magnetic flux differently from the MAD baseline and develops into a state that we identify as a Chimera MAD: it resembles the standard FM MAD in some diagnostics while remaining SANE-like or otherwise nonstandard in others. It reaches a MAD-level horizon magnetic flux and produces a high electromagnetic jet power during an extended quiescent high-flux interval with no clear eruptions, yet it remains distinct in density structure, mass-flow channels, force balance, angular-momentum transport, and outflow geometry even after eruptions begin.

We therefore use the Chimera comparison to ask which elements of the usual MAD phenomenology are most closely tied to horizon-scale magnetic flux and which depend on how that flux is supplied. Horizon fluxes and jet power measure the buildup of coherent flux and its coupling to the black hole; midplane morphology and time variability identify when that flux becomes eruptive; radial and meridional diagnostics show how the flow reorganizes its density, kinematics, force balance, and angular-momentum transport; and latitude-resolved mass fluxes show where the accretion channels actually form. In this sense, the MAD designation is treated as a set of related physical statements rather than as a single scalar threshold.

The rest of the paper explores these diagnostics quantitatively. In Section~\ref{sec:methods}, we describe our numerical methods and simulations. In Section~\ref{sec:fluid}, we compare fluid diagnostics and show that strong horizon-scale magnetization, efficient jet production, disk structure, eruption activity, and transport geometry are connected but separable. Finally, in Section~\ref{sec:discussion}, we summarize why the SANE/MAD distinction remains valuable but incomplete and note limitations and directions for further study.

\section{Numerical setup}
\label{sec:methods}

We analyze a set of three-dimensional general relativistic magnetohydrodynamics (GRMHD) simulations of radiatively inefficient accretion flows in the Kerr spacetime. The simulations are chosen to compare three ways of supplying gas and magnetic flux to the black hole: a weakly magnetized SANE flow, a standard MAD that rapidly reaches large horizon-threading flux and quickly begins to exhibit flux eruptions, and a Chimera MAD initialized to follow a different history of mass and coherent magnetic-flux supply. Our comparison asks how the supply of mass, angular momentum, and magnetic flux is related to the development of dynamically important magnetic flux and to the route to flux eruptions, and how that route is reflected in disk dynamics, force balance, angular momentum transport, and outflow energetics. In this sense the setup is a way to generate different accretion histories rather than the endpoint of the interpretation: different reservoirs and supply geometries need not erase into the same accretion state.

\subsection{General relativistic magnetohydrodynamics}

We use the GPU-accelerated code \athenak to solve the equations of ideal GRMHD and evolve the fluid. All simulations are performed in ingoing horizon-penetrating Kerr--Schild coordinates, and the spacetime is fixed according to the Kerr metric with spin parameter $\bhspin = J c / GM^2$, which we set to $\bhspin = 0.9$ for all simulations considered in this work. The GRMHD equations are
\begin{align}
\partial_t \left( \sqrt{-g} \rho u^t \right) &= -\partial_i \left( \sqrt{-g} \rho u^i \right), \label{eqn:massConservation}\\
    \partial_t \left( \sqrt{-g} {T^t}_{\nu} \right) &= - \partial_i \left( \sqrt{-g} {T^i}_{\nu} \right) + \sqrt{-g} {T^{\kappa}}_{\lambda} {\Gamma^{\lambda}}_{\nu\kappa},  \\
\partial_t \left( \sqrt{-g} B^i \right) &= - \partial_j \left[ \sqrt{-g} \left( b^i u^j - b^j u^i \right) \right], \label{eqn:fluxConservation}
\end{align}
along with the no-monopoles constraint
\begin{align}
\partial_i \left( \sqrt{-g} B^i \right) &= 0. \label{eqn:monopoleConstraint}
\end{align}
Here, $\rho$ is the rest-mass density, $u^\mu$ is the fluid four-velocity, $b^\mu$ is the magnetic four-vector for the field in the fluid frame, and $B^i = \;^\star F^{it}$. The determinant of the metric is $g$, and ${\Gamma^\lambda}_{\mu\nu}$ are the Christoffel symbols of the background spacetime. The total stress-energy tensor is given by
\begin{align}
{T^{\mu\nu}} &= \left( \rho + u + P_{\rm gas} + b^2\right)u^{\mu}u^{\nu} \nonumber  \\
&\qquad \quad + \left(P_{\rm gas} + \dfrac{b^2}{2} \right)g^{\mu\nu} - b^{\mu}b^{\nu},
\label{eqn:mhdTensor}
\end{align}
where $u$ is the internal energy, $b^2 = b^\lambda b_\lambda$, and the fluid pressure is obtained via the ideal gas law $P_{\rm gas} = (\hat{\gamma} - 1) u$ with adiabatic index $\hat{\gamma}$. In our simulations, we set $\hat{\gamma} = 13/9$ for the single-fluid dynamical evolution, which is appropriate for a plasma with the same ion and electron temperatures but relativistic electrons and non-relativistic ions (see also \citealt{gammie_2025_adiabaticindex}). The equations of ideal GRMHD are invariant under rescalings of both length and mass/energy density, which is appropriate for the low-accretion-rate advection dominated/radiatively inefficient accretion flows (RIAFs) that are believed to characterize a significant fraction of nearby galaxies \citep{ho_2008_llagn}.

\athenak solves these equations on a logically Cartesian mesh in Cartesian Kerr--Schild coordinates with static mesh refinement, concentrating resolution at small radii near the black hole where dynamical timescales are shortest and the flow is most variable. In practice, the coarsest refinement grid covers a cube centered on the black hole, and each successive refinement level spans a concentric cube with side lengths half those of its parent. The magnetohydrodynamics sector is evolved using constrained transport \citep{evans_1988_ct} with upwinded electric fields computed following \citet{gardiner_2005_ct,gardiner_2008_ct}. We use the first-order flux correction fallback as implemented in \athenak \citep{lemaster_2009_fofc,stone_2026_athenak} in the rare zones where the higher-order update would otherwise produce negative density or pressure. Further details about \athenak and the simulation toolkit can be found in \citet{wong_2022_patoka,wong_2025_mixing,stone_2026_athenak}.

\subsection{Simulations overview}

Each simulation is initialized with a hydrodynamic equilibrium torus and a specified magnetic vector potential. We seed the fluid with small perturbations in the thermal energy to promote the development of the magnetorotational instability \citep[MRI;][]{balbus_1991_mri}, which then drives turbulent angular-momentum transport and allows accretion to begin. The standard SANE and MAD simulations are the $\bhspin=0.9$ high-resolution FM tori studied in \citet{wong_2025_mixing}, with an inner edge at $20\,GM/c^2$ and a pressure maximum at $41\,GM/c^2$. They provide a familiar baseline: the gas initial condition is the same in the two runs, while the magnetic-field prescription determines whether the flow remains SANE-like or rapidly accumulates enough large-scale flux to become a standard MAD. Such torus initial conditions have often been used to model advection-dominated accretion flows onto black holes like M87$^*$ and \sgra, especially when simulations are compared with observations \citep[e.g.,][]{porth_2019_sanecomp,wong_2022_patoka,eht_m87_5,eht_m87_8,eht_sgra_5,eht_sgra_8,narayan_2022_jetsurvey,dhruv_2025_surveyv3,chael_2025_radsurvey}.

To generate the Chimera accretion history, we use the equilibrium torus solution of \citet{chakrabarti_1985_thickdisk} with an inner edge $r_{\rm in}=15$ and a pressure maximum $r_{\rm peak}=58$. This choice provides a more extended reservoir with a prescribed radial angular momentum gradient rather than the constant-angular-momentum structure of the standard FM torus. In the context of our work, the Chakrabarti torus modifies the reservoir of mass and angular momentum through which magnetic flux is advected toward the black hole and therefore changes the history by which the hole is fed. The resulting flow therefore probes how changes in mass, angular momentum, and magnetic-flux supply alter the route to strong horizon-scale magnetization.

\begin{figure*}[th!]
\centering
\includegraphics[width=\linewidth]{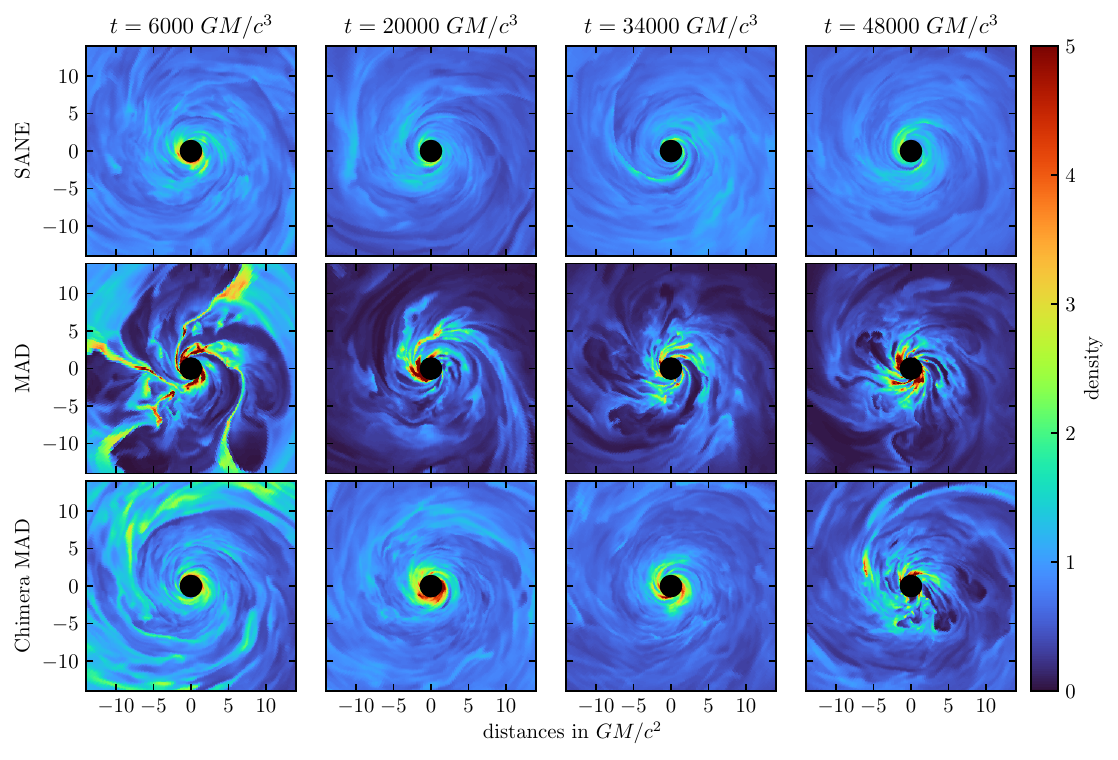}
\caption{
Representative midplane slices of fluid-frame density for the SANE, standard MAD, and Chimera MAD simulations, with columns showing snapshots evenly spaced in time. The images are normalized so that the mean density at $r=4\,GM/c^2$ is unity. The black circle marks the event horizon. The SANE flow remains comparatively smooth, the standard MAD shows strong time variability and low-density inner cavities, and the Chimera MAD resembles the smoother SANE morphology for longer before developing more pronounced low-density structures at late times.
}
\label{fig:midplane_splash}
\end{figure*}

The magnetic field is specified through the vector potential $A_\phi$, which is then differenced on the constrained-transport grid so that the numerical magnetic field satisfies the no-monopoles constraint to machine precision. After constructing the field, we rescale it so that the ratio of the maximum gas pressure in the torus to the maximum magnetic pressure is $100$. For the FM SANE and MAD baselines, we use the standard single-polarity loop prescriptions
{\small
\begin{align}
A_{\phi;\, \rm SANE} &= \max \left[ \dfrac{\rho}{\rho_{\max}} - 0.2, 0 \right] \\
A_{\phi;\, \rm MAD} &= \max \left[ \dfrac{\rho}{\rho_{\max}} \left( \dfrac{r}{r_{\rm in}} \sin\theta\right)^3 e^{-r/400} - 0.2, 0 \right],
\end{align}
}
where $r_{\rm in}=20$ is the inner edge of the initial torus. For the Chakrabarti torus, however, the standard FM MAD vector potential does not deliver enough coherent magnetic flux to the event horizon to produce the behavior studied here. We therefore thread the Chimera reservoir with a larger-scale vertical field and set
{\small
\begin{align}
A_{\phi;\,{\rm Chimera}} = \left(\dfrac{r_{\rm cyl}}{r_{\rm in}}\right)^2 e^{-r_{\rm cyl}/r_{\rm falloff}} - e^{-r_{\rm in}/r_{\rm falloff}},
\end{align}} \\
where $r_{\rm cyl}=r\sin\theta$, $r_{\rm in}=15$, and $r_{\rm falloff}=80$. The vector potential is set to zero for $r_{\rm cyl}<r_{\rm in}$. This construction is deliberately different from the standard FM MAD loop and gives the extended Chakrabarti torus access to a coherent vertical field reservoir.

\begin{table*}[t]
\centering
\caption{Summary of the simulations analyzed in this work. The SANE and standard MAD models are the $\bhspin=0.9$ high-resolution simulations from \citet{wong_2025_mixing}. The Chimera MAD uses the same spin and numerical resolution as the FM MADs, but the initial torus is threaded by the large-scale vertical field described in the text.}
\label{tab:simulations}
\begin{tabular}{lllll}
\hline
Model & Fluid torus & Magnetic field & Analysis interval & Role in comparison \\
\hline
SANE & Fishbone--Moncrief & $A_{\phi;\,{\rm SANE}}$ & $20000$--$50000\,GM/c^3$ & weakly magnetized baseline \\
MAD & Fishbone--Moncrief & $A_{\phi;\,{\rm MAD}}$ & $20000$--$50000\,GM/c^3$ & baseline with rapid-onset eruptions \\
Chimera MAD (early) & Chakrabarti & vertical $A_{\phi;\,{\rm Chimera}}$ & $20000$--$40000\,GM/c^3$ & delayed, comparatively steady \\
Chimera MAD (late) & Chakrabarti & vertical $A_{\phi;\,{\rm Chimera}}$ & $40000$--$50000\,GM/c^3$ & late, more eruptive \\
\hline
\end{tabular}
\end{table*}

Our simulation domain extends over a cube reaching $\pm 1024\ GM/c^2$ in each Cartesian Kerr--Schild coordinate direction. Outflow boundary conditions are imposed at the domain edges, while the central black hole is treated with an excision boundary \citep{olivares_2019_bhac,ressler_2021_sphericalaccretion,stone_2026_athenak}. All production runs use the same nested static mesh refinement grid as the high-resolution simulations in \citet{wong_2025_mixing}. The coarsest level covers the full domain, each successive level covers a concentric cube with half the side length of its parent, and the finest region spans the innermost $\pm 8\ GM/c^2$ with 16 cells per $GM/c^2$. We use this fiducial 16/8 resolution for all figures so that the SANE, standard MAD, and Chimera MAD comparisons are resolution matched. As a resolution check, we also evolved a Chimera run at twice the linear resolution over the same time interval and found quantitatively consistent diagnostics. We evolve each simulation until $t = 50000\,GM/c^3$.

\section{Fluid diagnostics}
\label{sec:fluid}

We now compare the dynamical and kinematic properties of the SANE, standard MAD, and Chimera MAD simulations to determine how accreting plasma reaches the black hole and how magnetic fields reorganize the flow. The diagnostics connect time-dependent horizon fluxes and midplane morphology with spatial flow structure, radial force balance, angular-momentum transport, and jet/outflow energetics. Unless otherwise stated, we analyze the SANE and standard MAD simulations over $20000$--$50000\,GM/c^3$, after the initial transient has passed. For the Chimera MAD, we split the analysis into an early extended high-magnetization interval between $20000$--$40000\,GM/c^3$, during which time flux eruptions do not occur, and a late interval, $40000$--$50000\,GM/c^3$, after low-density bubbles and sharp decreases in the horizon-scale magnetic flux begin. 

\begin{figure*}[th!]
\centering
\includegraphics[width=\linewidth]{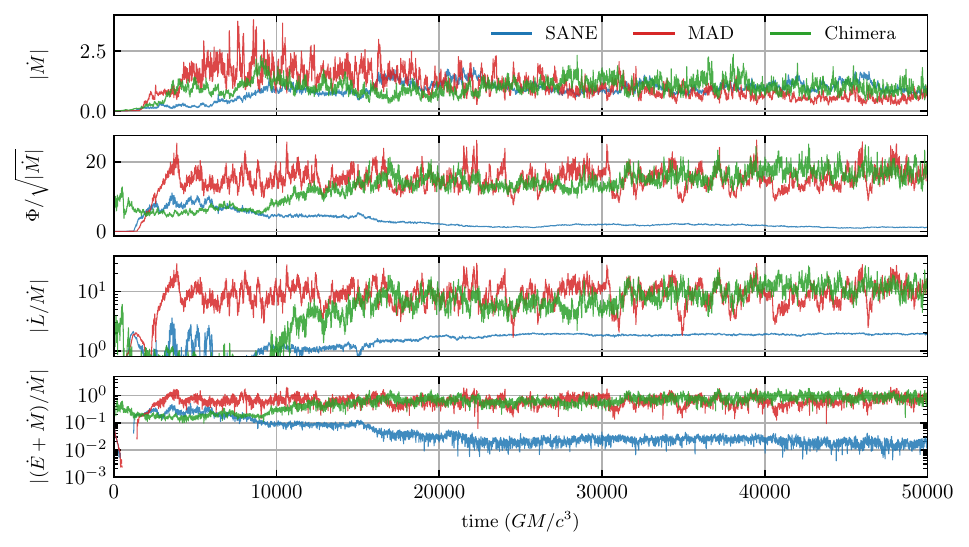}
\caption{
Time series of horizon fluxes for the three simulations. From top to bottom, the panels show the mass accretion rate normalized by its post-transient mean, as well as the instantaneous dimensionless horizon-threading magnetic flux $\phi_{\rm BH}=\Phi_{\rm BH}/|\dot M|^{1/2}$, specific angular-momentum flux $|\dot L/\dot M|$, and rest-mass-subtracted specific energy flux $|(\dot E+\dot M)/\dot M|$. The standard MAD rapidly reaches a large horizon-threading magnetic flux and begins to erupt, while the SANE remains weakly magnetized after its initial transient. The magnetic flux in the Chimera MAD approaches saturation in $\phi$ more gradually and remains at large flux for an extended interval before entering its eruptive phase at $t \approx 40,\!000\,GM/c^3$.
}
\label{fig:horizon_fluxes}
\end{figure*}

\subsection{Temporal variability}
\label{sec:fluid_timeseries}

We begin by comparing the time-dependent fluid structure of the three accretion states. Figure~\ref{fig:midplane_splash} shows representative midplane density slices from the SANE, standard MAD, and Chimera MAD simulations. The SANE flow remains comparatively smooth throughout the simulation: its inner disk is turbulent and nonaxisymmetric, but the density field is organized primarily into the spiral features of a weakly magnetized accretion flow. The standard MAD is qualitatively different: once sufficient large-scale magnetic flux accumulates near the black hole, the inner flow becomes strongly nonaxisymmetric, and dense accretion streams coexist with low-density, magnetically dominated cavities in the disk body. Table~\ref{tab:simulations} summarizes the set of simulations we consider.

The Chimera MAD follows a different evolution. Despite its relatively strong, coherent magnetization, its early-time midplane morphology resembles the SANE flow: the disk is turbulent, and the inner flow is not yet dominated by magnetic bubbles. As magnetic flux accumulates, the inner disk becomes increasingly magnetically structured. By late times, the Chimera MAD develops clear low-density bubbles and strongly nonaxisymmetric inner-flow morphology near the event horizon, indicating that the flow has entered an eruptive phase. Notably, even though the Chimera MAD quickly develops a large magnetic flux near the event horizon, it only begins to erupt after an extended interval in which its disk structure and fluid variability are closer to the steadier SANE flow than to the standard MAD.

We quantify the accretion flow evolution by evaluating fluxes through the event horizon. We define
\begin{align}
\dot{M}
  &= \int_0^{2\pi}\int_0^\pi
      \left(-\rho u^r\right)\sqrt{-g}\,d\theta\,d\phi, \\
\Phi_{\rm BH}
  &= \frac{1}{2}\int_0^{2\pi}\int_0^\pi
      \left|B^r\right|\sqrt{-g}\,d\theta\,d\phi, \\
\dot{L}
  &= \int_0^{2\pi}\int_0^\pi
      T^r{}_\phi\sqrt{-g}\,d\theta\,d\phi, \\
\dot{E}
  &= \int_0^{2\pi}\int_0^\pi
      \left(-T^r{}_t\right)\sqrt{-g}\,d\theta\,d\phi .
\end{align}
\noindent Here the integrals are taken over the full angular shell. The sign convention is chosen so that $\dot M>0$ corresponds to inward rest-mass accretion, while positive $\dot L$ and $\dot E$ correspond to outward fluxes of angular momentum and energy. The factor of $1/2$ in $\Phi_{\rm BH}$ translates the total unsigned magnetic flux through the horizon to an equivalent one-hemisphere flux. The angular-momentum and energy fluxes are the radial components of the conserved currents associated with the axial and time-translation Killing vector fields, $\xi^\mu_{(\phi)} \equiv \partial_\phi$ and $\xi^\mu_{(t)} \equiv \partial_t$, respectively.

Figure~\ref{fig:horizon_fluxes} shows the evolution of these horizon fluxes over the full simulation durations. We normalize the mass accretion rate $\dot{M}$ so that its average value for $t > 10000\,GM/c^3$ is unity to compare variability between the simulations and because ideal GRMHD is scale-free, making the absolute accretion rate more challenging to interpret.\footnote{When starting from the same Fishbone--Moncrief fluid initial torus, the standard MAD accretes more rapidly than the standard SANE in code units.} A standard quantitative hallmark of MAD accretion is instead the growth of the normalized horizon-threading magnetic flux until it reaches a saturation level, $\phi_{\rm BH}\equiv \Phi_{\rm BH}/\sqrt{|\dot{M}|}\sim 15$ \citep[e.g.,][and see also \citealt{narayan_2022_jetsurvey,dhruv_2025_surveyv3} for the dependence on black hole spin]{tchekhovskoy_2011_mad}.\footnote{We calculate magnetic field strengths in Lorentz--Heaviside units. In Gaussian units, which differ by a factor of $\sqrt{4\pi}$, the saturation value is $\phi \approx 50$.} In addition to this dimensionless magnetic flux $\phi$, we also consider the angular-momentum flux $|\dot{L}/\dot{M}|$ and the rest-mass-subtracted energy flux $|(\dot{E}+\dot{M})/\dot{M}|$ normalized to the instantaneous mass accretion rate.

At early times, the mass accretion rate reflects the transient arrival of material from the initial torus before settling into a fluctuating quasi-steady state. The normalized flux $\phi_{\rm BH}$ is the more direct diagnostic of whether coherent magnetic flux has become dynamically important near the horizon. In the SANE model, $\phi_{\rm BH}$ remains small after the initial transient, and the flow remains far from the high-flux regime associated with the standard MAD model. The angular-momentum and energy fluxes are correspondingly weaker and less bursty than in the MAD models. This is the time-domain counterpart of the smooth midplane morphology in Figure~\ref{fig:midplane_splash}: the magnetic field participates in turbulent angular-momentum transport, but it does not accumulate enough coherent horizon-threading flux to reorganize the inner flow. The standard MAD reaches large $\phi_{\rm BH}$ rapidly. Once this occurs, the horizon flux fluctuates around a high value and undergoes repeated sharp decreases associated with flux-eruption episodes. These drops in $\phi_{\rm BH}$ are accompanied by strong variability in $\dot{M}$, $\dot{L}$, and $\dot{E}$, and the flow is regulated by the repeated formation and disruption of magnetically dominated bubbles in the inner disk.

The Chimera MAD follows a distinct evolution. Its horizon magnetic flux grows more slowly than in the standard MAD but still reaches a large dimensionless value $\approx 15$, comparable to the MAD saturation level.\footnote{In general, the value of the flux threshold for the onset of eruptive behavior likely depends on disk geometry and mass loading.} Nevertheless, over an extended interval, the horizon fluxes do not exhibit the same strongly bursty MAD behavior. Operationally, the late Chimera interval is identified by the appearance of sharp decreases in the unnormalized horizon flux and the simultaneous development of the low-density midplane bubbles seen in Figure~\ref{fig:midplane_splash}; the earlier interval is therefore not simply noise before saturation, but a long-lived high-flux, jet-capable state with weaker eruption phenomenology. The later eruptions also do not erase the history of the flow, since the eruptive Chimera interval remains distinct from the standard MAD in the diagnostics below.

\begin{figure*}[th!]
\centering
\includegraphics[width=.9\linewidth]{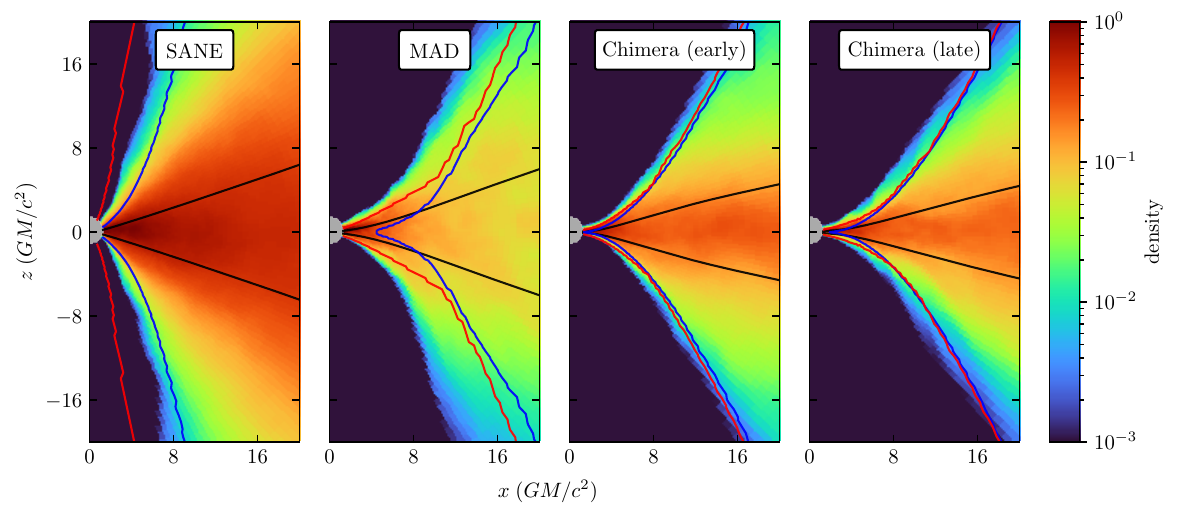}
\caption{
Time- and azimuth-averaged density structure for the three simulations. The red curve marks $(\beta\gamma)_\infty=1$, the blue curve marks $\sigma=1$, and the black curve shows the density scale height. The $\sigma=1$ contour also defines the mask used to exclude magnetically dominated zones from the disk-body radial profiles presented below. The SANE flow has a broad, matter-dominated disk and gradual transition to the polar region, while the standard MAD has a more evacuated funnel and thinner inner disk. The Chimera MAD changes between its early and late intervals, but its density distribution and funnel-wall geometry remain distinct from the standard MAD.
}
\label{fig:profiles2d}
\end{figure*}

\subsection{Meridional structure}

We next use time- and azimuth-averaged meridional profiles to compare the persistent disk, funnel, and disk--funnel geometry. Figure~\ref{fig:profiles2d} shows the averaged density in the meridional plane. The SANE flow is relatively thick and remains dense at larger radii away from the event horizon. It launches a comparatively narrow jet, but the transition between the disk, corona, and jet region is gradual. In the standard MAD, strong magnetic flux evacuates the polar region more effectively and compresses the inflowing gas. The Chimera MAD does not simply interpolate between the standard MAD and SANE states: at late times the flow becomes more MAD-like and the polar region is more strongly evacuated, leading to a somewhat thinner disk structure, but the density and funnel-wall geometry remain organized differently from the standard MAD.

The curves in Figure~\ref{fig:profiles2d} denote several characteristic surfaces in the flow. The blue contour marks the magnetization threshold $\sigma \equiv b^2/\rho = 1$, which approximately separates the dense, matter-dominated disk from the magnetically dominated funnel and coronal regions. The red contour marks $(\beta\gamma)_\infty = 1$.\footnote{Here $\beta$ is the familiar normalized velocity $\beta = v/c$ (rather than plasma $\beta$), which is used to construct this common diagnostic of relativistic outflow material. The quantity $(\beta\gamma)_\infty$ is closely related to Bernoulli-type jet selections \citep[e.g.,][]{eht_m87_5, dhruv_2025_surveyv3}. See Section~\ref{sec:jet_outflows} for more detail.} The black curves show the disk scale height, which we define from the vertically symmetric density distribution as the density-weighted angular distance from the midplane,
\begin{align}
    \frac{h}{r}
    \equiv
    \frac{\int \rho \, |\theta-\pi/2| \; \sqrt{-g}\, d\theta\, d\phi}
         {\int \rho \; \sqrt{-g}\, d\theta\, d\phi}.
\end{align}
This quantity provides a simple measure of the geometric thickness of the dense flow and allows a direct comparison of how puffed up or compressed the disk is in the different accretion states. As representative values, at $r=25\,GM/c^2$ we find $h/r\simeq0.3$ for both SANE and standard MAD and $h/r\simeq0.2$ for both Chimera epochs, while at $r=4\,GM/c^2$ the corresponding values are approximately $0.25$, $0.15$, $0.12$, and $0.10$ for the SANE, MAD, early Chimera, and late Chimera flows.

\begin{figure*}[th!]
\centering
\includegraphics[width=.9\linewidth]{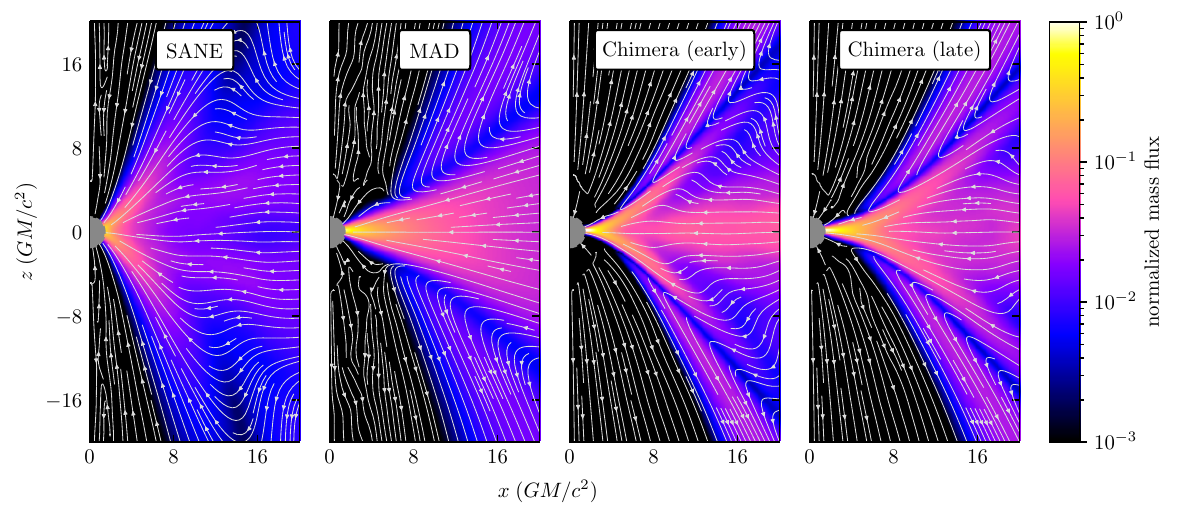}
\caption{
Time-averaged poloidal mass flux in the SANE, standard MAD, and Chimera MAD flows, normalized by the accretion rate through the event horizon. Streamlines trace the mean poloidal flow direction. The SANE flow has prominent off-midplane accretion channels, whereas the standard MAD and Chimera MAD show more organized inflow--outflow structure near the black hole. The Chimera MAD differs from the standard MAD in the detailed geometry of its funnel wall and surface-layer flow.
}
\label{fig:massflux}
\end{figure*}

\begin{figure*}[th!]
\centering
\includegraphics[width=.8\linewidth]{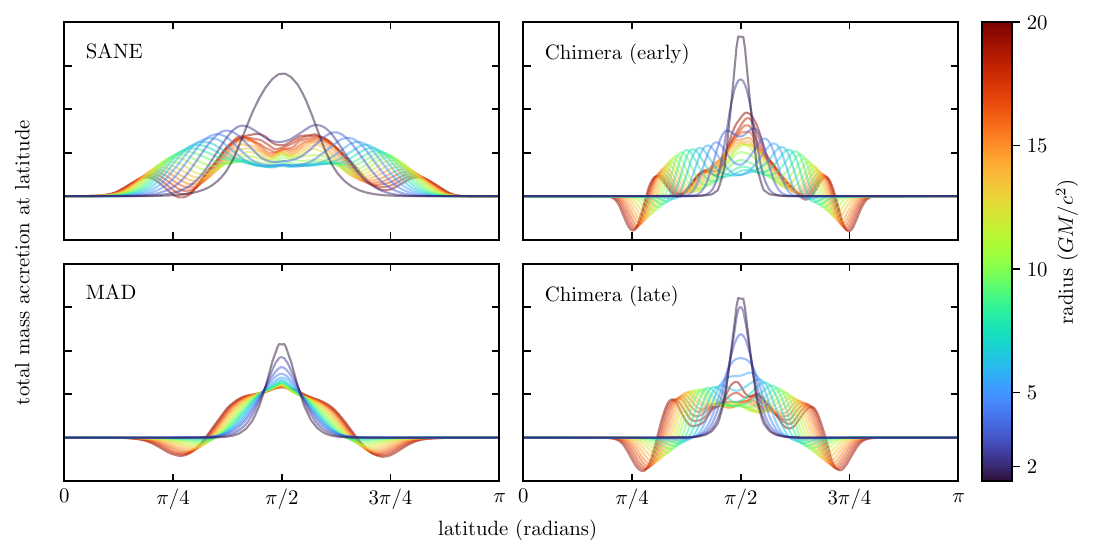}
\caption{
Radial mass accretion rate as a function of latitude for the SANE, standard MAD, and early and late Chimera MAD intervals. Positive values denote inward accretion and negative values denote outward mass flux. The SANE flow shows broad off-midplane accretion peaks, while the standard MAD places more of the inward mass flux near the midplane and develops outflowing regions at higher latitudes. The Chimera MAD evolves from a more vertically structured early pattern toward a sharper inner inflow component at late times.
}
\label{fig:massflux_latitude}
\end{figure*}

The mass-flux map in Figure~\ref{fig:massflux} shows the time-averaged poloidal mass flux, with streamlines showing the direction of the mass flow from $\langle \rho u^{i}\sqrt{-g}\rangle$. Accretion is not purely equatorial. In the SANE flow, a significant fraction of the mass is routed through off-midplane layers above and below the dense midplane; the mass flux peaks in channels along the disk/funnel interface, where the radial velocity is largest, so the total mass flux can peak there even though the density is lower. Figure~\ref{fig:massflux_latitude} quantifies this statement by decomposing the radial accretion rate into its latitudinal contributions at different radii. In the SANE flow, the inward contribution is broad and peaks away from the midplane over a wide range of radii, confirming that the off-midplane channels carry a substantial part of the accretion flow. The standard MAD is more disk-centered: matter flows inward through a broad body of the disk, while winds are launched from the surface. The Chimera MAD combines features of both baselines, with surface winds but, especially at early times, accretion peaking along channels away from the midplane.

\begin{figure*}[th!]
\centering
\includegraphics[width=0.8\linewidth]{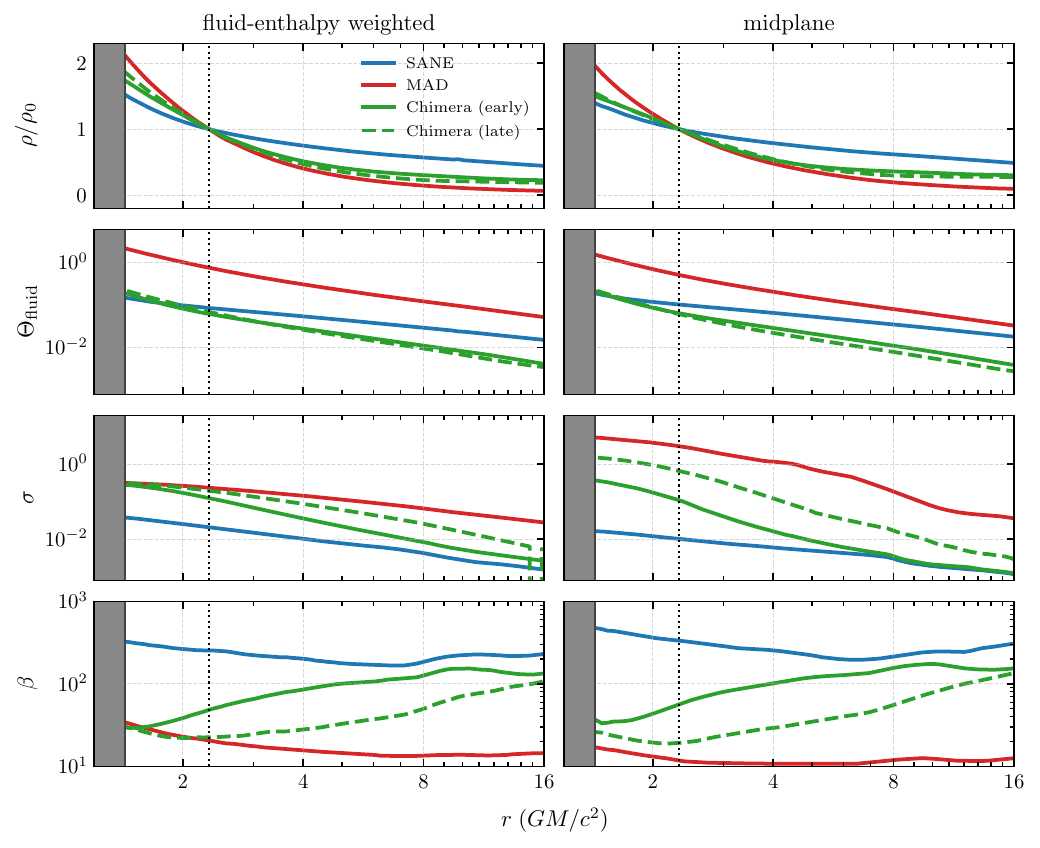}
\caption{
Radial profiles of basic fluid diagnostics for the SANE, standard MAD, and Chimera MAD flows, shown as disk-body averages (left) and midplane values (right). From top to bottom, the panels show density normalized to its value at the innermost stable circular orbit (ISCO), dimensionless fluid temperature $\Theta_{\rm fluid}$, magnetization $\sigma \equiv b^2/\rho$, and plasma $\beta \equiv 2P_{\rm gas}/b^2$. Comparing the two columns highlights vertical stratification, with the midplane more strongly magnetized than the disk, especially in the MAD and late Chimera states. The gray shaded region marks the event horizon and the dotted vertical line marks the ISCO. The SANE flow remains denser and less magnetized, the standard MAD is hotter and more strongly magnetized, and the Chimera MAD evolves toward lower density and higher magnetization while remaining distinct from the standard MAD.
}
\label{fig:radial_profiles}
\end{figure*}

The off-midplane SANE channels are reminiscent of a broader class of vertically structured accretion phenomena in which the density maximum and the dominant transport channel need not coincide. In classical magnetic-flux advection models, turbulent diffusion makes it difficult for a thin disk to drag in weak poloidal flux \citep{lubow_1994_fielddragging}, motivating models in which surface layers or coronal flows carry field inward more efficiently \citep{rothstein_2008_fieldadvection,lovelace_2009_fieldadvection}. Global GRMHD simulations have also identified a related ``coronal mechanism'' in which large-scale poloidal flux is transported largely outside the dense disk body \citep{beckwith_2009_poloidalflux}. \citet{guilet_2012_fluxtransport,guilet_2013_fluxtransport} showed that vertical structure can substantially modify the relative transport of mass and magnetic flux, in part because low-density regions away from the midplane can have larger radial velocities. The present SANE flow gives a complementary mass-transport view in a thick GRMHD RIAF: the dominant inward mass flux can be displaced away from the midplane and concentrated near the disk/funnel interface, where the radial velocity is larger despite the lower density.

\subsection{Radial profiles}

\begin{figure}[th!]
\centering
\includegraphics[width=0.9\linewidth]{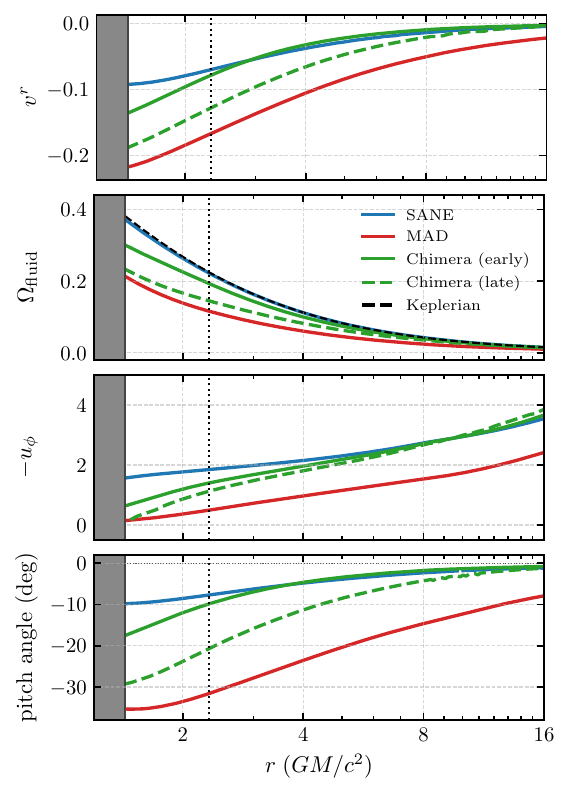}
\caption{
Radial profiles of kinematic diagnostics for the SANE, standard MAD, and early and late Chimera MAD intervals. The panels show radial velocity, fluid angular velocity, specific angular momentum $-u_\phi$, and fluid-motion pitch angle $\arctan(v^r/R\Omega)$. The shaded region extends to the event horizon and the dashed line marks the ISCO. The SANE flow accretes more slowly and rotates closer to the Keplerian angular velocity, while the standard MAD is more rapidly plunging, more sub-Keplerian, and has a more negative pitch angle. The Chimera MAD shifts from a more SANE-like early profile toward a more MAD-like late profile.
}
\label{fig:velocity_profiles}
\end{figure}

We next compress the axisymmetric structure into radial profiles of the disk body and midplane to provide a compact comparison of the thermodynamic, magnetization, and kinematic state of the gas. Since the midplane profile does not necessarily represent the state of the flow or characterize the predominant motion of the accreting matter (e.g., Figure~\ref{fig:massflux}), we also often report fluid-enthalpy-weighted angular profiles restricted to the matter-dominated portion of the flow
\begin{align}
    \left\langle q \right\rangle_{w_{\rm fl},\,\sigma<1}
    =
    \dfrac{\int q \; \mathcal{M} \;  w_{\rm fl}\,u^t  \sqrt{-g}  \, d\theta \, d\phi}{\int \; \mathcal{M} w_{\rm fl}\,u^t \sqrt{-g} \, d\theta \, d\phi},
\end{align}
where the mask $\mathcal{M}$ is unity in zones with $\sigma<1$ and zero otherwise. Here $w_{\rm fl} = \rho + u + P_{\rm gas}$ is the enthalpy of the fluid, which we use so that hotter, pressure-supported material contributes consistently to disk-body averages. The factor $u^t\sqrt{-g}$ converts the weighting to the natural coordinate-time volume measure on a Kerr--Schild time slice.

Figure~\ref{fig:radial_profiles} shows the density, dimensionless fluid temperature $\Theta = P_{\rm gas} / \rho$, magnetization $\sigma = b^2/\rho$, and plasma $\beta$. The SANE flow remains the densest and least magnetized of the three states over most of the inner disk, with large plasma beta and low magnetization. Magnetic fields participate in turbulent transport in this state, but they do not dominate the mean force balance or global structure. The standard MAD is hotter, more rarefied, and more strongly magnetized. Its lower $\beta$ and higher $\sigma$ reflect the accumulation of dynamically important large-scale magnetic flux near the black hole.

The Chimera MAD evolves systematically across its two intervals rather than collapsing onto either limiting baseline. At early times, its density and magnetization profiles remain closer to SANE, especially outside the immediate near-horizon region. At late times, the density decreases and the magnetization increases, moving the profile toward the standard MAD state. The transition is not simply a uniform rescaling of the density or pressure. The late-time Chimera MAD therefore occupies an intermediate but distinct region of parameter space: it is much more strongly magnetized than the SANE, but it does not become identical to the standard MAD in either density or $\beta$.

The kinematic profiles in Figure~\ref{fig:velocity_profiles} show the corresponding change in orbital state. Plotted are the radial velocity of the flow $v^r = u^r/u^t$, the fluid angular velocity $\Omega = u^\phi/u^t$, the specific angular momentum $-u_\phi$, and the fluid-motion pitch angle $\arctan(v^r/R\Omega)$, where $R$ is the cylindrical radius. For reference, we compare the angular velocities to the prograde circular-geodesic ``Keplerian'' angular velocity, $\Omega_{\rm Kep}=\left(r^{3/2}+\bhspin\right)^{-1}$. The SANE flow remains comparatively close to the Keplerian angular velocity while the standard MAD rotates substantially below the Keplerian rate, has a larger inward radial velocity, and a correspondingly more negative pitch angle throughout the inner disk. This is one basic dynamical signature of strong magnetic regulation: angular momentum has been removed, and the plasma no longer relies primarily on orbital motion for radial support.

The Chimera MAD evolves in the direction expected for a more strongly magnetized flow without becoming a duplicate of the standard MAD. At early times, its angular velocity, radial velocity, and pitch angle lie between the SANE and standard MAD profiles. At late times, the flow begins to exhibit more rapid infall, becomes more sub-Keplerian, and develops a more plunging pitch angle, approaching the MAD solution over much of the inner disk. The disk changes its orbital dynamics as magnetic flux accumulates and the inner flow becomes more magnetized.

\subsection{Force balance in the flow}

Figures~\ref{fig:radial_profiles} and \ref{fig:velocity_profiles} show that the flows differ substantially in their orbital kinematics: the MAD models rotate more slowly and fall inward more rapidly than the SANE flow, while the Chimera MAD evolves toward this more plunging configuration. The SANE/MAD distinction is often framed through horizon flux and variability; however, force balance provides a complementary local measure of how magnetic fields support and redirect the gas, so we now ask what stresses support or accelerate gas with these kinematics. In a weakly magnetized, nearly Keplerian disk, radial balance is set primarily by orbital motion, with pressure forces providing a correction. When magnetic forces become dynamically important, this separation can fail: sub-Keplerian gas requires additional radial support from gas pressure, magnetic pressure, or magnetic tension unless it is undergoing rapid inward acceleration. The force decomposition below measures the local non-geodesic acceleration produced by each of these stresses and shows how the radial support budget changes between the SANE, MAD, and Chimera MAD flows. We compute the force and acceleration at each location in each simulation snapshot before averaging in time or azimuth; the weights and masks used in the averages are likewise applied to the instantaneous fields.

We frame this as a local physical question: for a fluid element moving along a geodesic, what non-gravitational accelerations does the element experience in the radial and latitudinal directions? Starting from conservation of the stress-energy tensor, $\nabla_\mu T^{\mu\nu} = 0 $, the covariant force equation is obtained by projecting orthogonal to the fluid four-velocity. The spatial projection tensor is
\begin{align}
    h_{\mu\nu} = g_{\mu\nu}+u_\mu u_\nu ,
\end{align}
which satisfies $h^\mu{}_\nu u^\nu=0$. The projected equation of motion can be written schematically as
\begin{equation}
    \mathcal{M}_{\mu\nu} a^\nu
    =
    f^{\rm gas}_\mu
    +
    f^{\rm magP}_\mu
    +
    f^{\rm tens}_\mu ,
    \label{eq:force_balance_general}
\end{equation}
where
\begin{equation}
    a^\mu \equiv u^\alpha \nabla_\alpha u^\mu
\end{equation}
is the non-geodesic four-acceleration of the fluid, and the tensor $\mathcal{M}_{\mu\nu}$ is the effective inertia of the magnetized fluid projected into the local rest frame,
\begin{equation}
    \mathcal{M}_{\mu\nu} = \left(w_{\rm fl} +b^2\right)h_{\mu\nu} - b_\mu b_\nu .
    \label{eq:anisotropic_inertia}
\end{equation}
Here, the second term reflects the anisotropic inertia associated with magnetic tension: acceleration along the field and acceleration perpendicular to the field do not couple to the same effective inertia.

\begin{figure*}[th!]
\centering
\includegraphics[width=.92\textwidth]{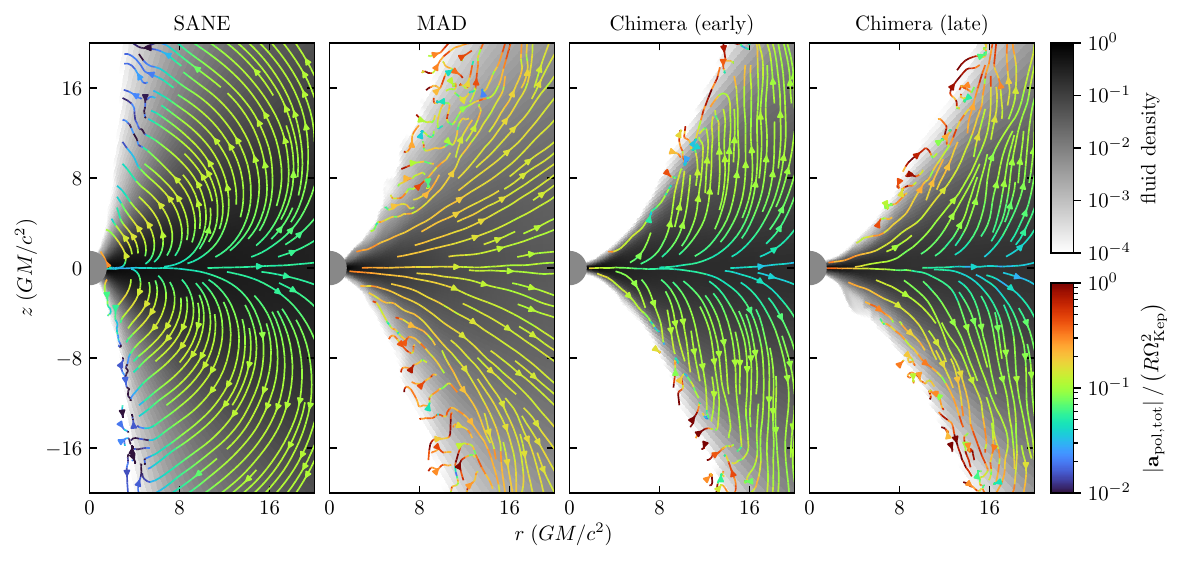}
\caption{
Time- and azimuth-averaged maps of the total poloidal non-geodesic acceleration, which measures how pressure and magnetic stresses support or redirect gas away from geodesic motion. Grayscale shows density; curves follow the acceleration direction (not mass-flow streamlines) and are colored by $|\mathbf{a}_{\rm pol,tot}|/(R\,\Omega_{\rm Kep}^2)$. The SANE flow has a smoother, disk-centered support field, while the standard MAD develops stronger acceleration in magnetized surface and funnel-wall layers. The Chimera MAD shifts toward this MAD-like force geometry at late times but retains a different disk/surface-layer pattern.
}
\label{fig:acceleration_2d_all}
\end{figure*}

\begin{figure*}[th!]
\centering
\includegraphics[width=.92\textwidth]{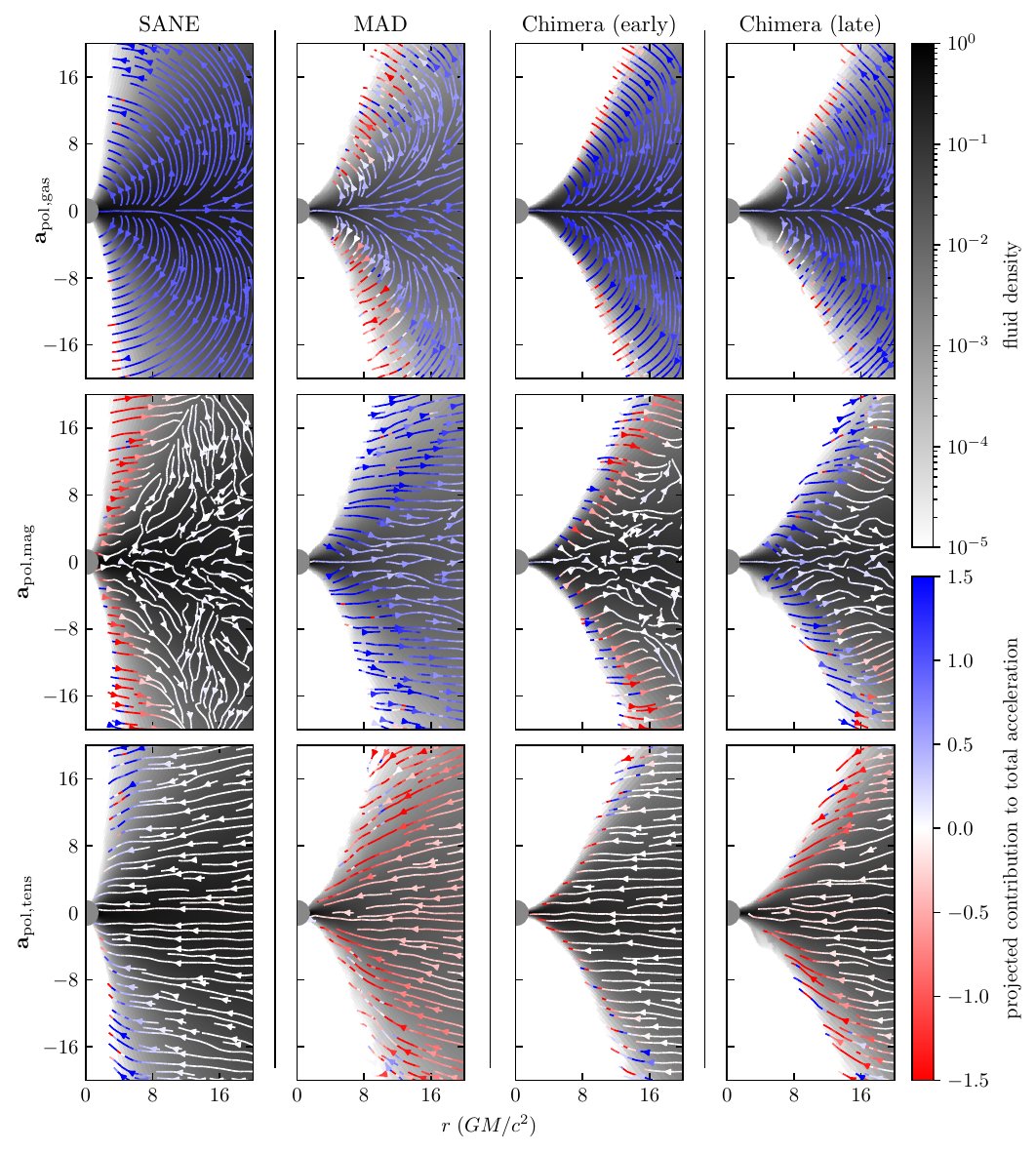}
\caption{
Decomposition of the total acceleration in Figure~\ref{fig:acceleration_2d_all}. Rows show the acceleration driven by gas-pressure gradients, magnetic-pressure gradients, and magnetic tension. Curves follow the direction of each component, while colors show the signed projection onto the total poloidal acceleration: blue supports the net acceleration, red opposes it, and white is weak or nearly orthogonal. The SANE non-geodesic acceleration is mostly gas-pressure driven, while the MAD states require magnetic pressure and tension in the inner disk, surface layers, and funnel wall. The Chimera MAD becomes more magnetically regulated at late times without reproducing the standard-MAD component geometry.
}
\label{fig:acceleration_2d_components}
\end{figure*}

\begin{figure}[th!]
\centering
\includegraphics[width=\linewidth]{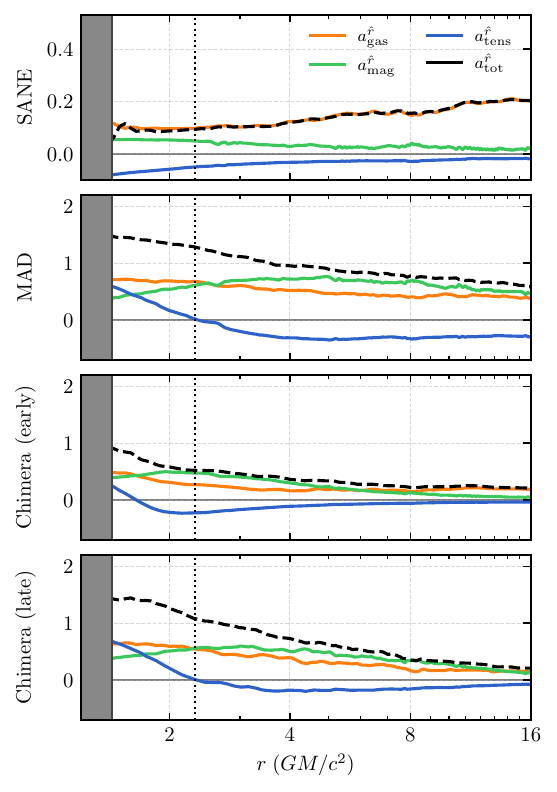}
\caption{
Radial profiles of the gas-pressure, magnetic-pressure, magnetic-tension, and total radial acceleration contributions for the SANE, standard MAD, and Chimera MAD flows. Accelerations are normalized by $R\,\Omega_{\rm Kep}^2$; positive values indicate outward acceleration and negative values assist infall. The gray shaded region marks the event horizon and the dotted vertical line marks the ISCO. The SANE flow is primarily gas-pressure supported, while the MAD and late Chimera MAD flows have magnetic-pressure and tension terms comparable to the gas-pressure contribution, indicating magnetic regulation of the inner-flow force balance.
}
\label{fig:force_decomp}
\end{figure}

The three terms on the right-hand side of Equation~\eqref{eq:force_balance_general} are
\begin{align}
    f^{\rm gas}_\mu
    &=
    - h_\mu{}^\nu \nabla_\nu P_{\rm gas} ,
    \\
    f^{\rm magP}_\mu
    &=
    - h_\mu{}^\nu \nabla_\nu \left(\frac{b^2}{2}\right) ,
    \\
    f^{\rm tens}_\mu
    &=
    h_\mu{}^\nu b^\alpha \nabla_\alpha b_\nu ,
\end{align}
and they correspond to the gas pressure force, the magnetic pressure force, and the field-line tension force, respectively. The first two terms measure pressure gradients in the instantaneous fluid rest frame. The third term measures the projected derivative of the magnetic field along itself and is therefore sensitive to the curvature and stretching of magnetic field lines.\footnote{Appendix~\ref{sec:force_identity_appendix} shows how the magnetic field divergence is absorbed into the anisotropic inertia tensor.}

Equation~\eqref{eq:force_balance_general} can be solved for the acceleration induced by each term separately so that
\begin{align}
    a^\mu_{\rm gas}
    &=
    \mathcal{N}^{\mu\nu}f^{\rm gas}_\nu , \\
    a^\mu_{\rm magP}
    &=
    \mathcal{N}^{\mu\nu}f^{\rm magP}_\nu , \\
    a^\mu_{\rm tens}
    &=
    \mathcal{N}^{\mu\nu}f^{\rm tens}_\nu ,
    \label{eq:component_accelerations}
\end{align}
where the inverse of the inertia tensor is
\begin{align}
    \mathcal{N}^{\mu\nu}
    =
    \frac{h^{\mu\nu}}{w_{\rm fl}+b^2}
    +
    \frac{b^\mu b^\nu}{w_{\rm fl}\,(w_{\rm fl}+b^2)} .
    \label{eq:inverse_inertia}
\end{align}

Coordinate components like $a^r$ are not ideal for interpreting the acceleration structure, because the coordinate basis vectors $\partial_r$ and $\partial_\theta$ are not spatial in the local fluid frame. We therefore project the acceleration onto a local orthonormal tetrad defined by the fluid motion (see Appendix~\ref{sec:coordinate_basis}). The hatted radial direction is constructed by projecting the coordinate direction $\partial_r$ into the instantaneous fluid rest space and normalizing it, so positive $a^{\hat r}$ points toward increasing Kerr-Schild radius as measured in the local comoving frame. In this frame, the radial force decomposition measures the additional non-geodesic radial acceleration required to support or accelerate gas with its instantaneous orbital state. The acceleration $a^{\hat r}$ thus corresponds to physical radial acceleration measured in a locally orthonormal frame tied to the fluid, and similarly for $a^{\hat{\theta}}$.

We compare each force component with the net acceleration by projecting the poloidal contribution onto the total poloidal acceleration. After converting to the plotted orthonormal meridional components, we define
\begin{align}
\mathcal{P}_i
\equiv
\frac{
\mathbf{a}_{\rm pol}^{\,i}\cdot \mathbf{a}_{\rm pol}^{\,\rm tot}
}{
|\mathbf{a}_{\rm pol}^{\,\rm tot}|^2
},
\end{align}
where $i=\{\mathrm{gas},\mathrm{magP},\mathrm{tens}\}$. Thus $\mathcal{P}_i>0$ denotes that component contributes along the direction of the net poloidal acceleration.
Note that this quantity is a directional contribution rather than a support fraction: if the net poloidal acceleration points inward, a positive $\mathcal{P}_i$ means that the component assists the inward acceleration.

Figure~\ref{fig:acceleration_2d_all} shows the total non-geodesic acceleration of fluid elements for each accretion state, and Figure~\ref{fig:acceleration_2d_components} decomposes that acceleration into gas-pressure, magnetic-pressure, and tension contributions. The total acceleration is normalized by $R\,\Omega_{\rm Kep}^2$, which is useful as an approximate measure of the acceleration compared to orbital support: values near unity indicate that the non-geodesic acceleration is dynamically comparable to the characteristic radial acceleration scale of the orbit. We use this only as a characteristic scale; it is not the exact criterion for centrifugal support, especially away from the midplane, near the horizon, and inside the innermost stable circular orbit (ISCO). In the SANE flow, the poloidal acceleration field is comparatively smooth and remains concentrated in the dense disk and funnel-wall region. The projected contributions show that this acceleration is primarily gas-pressure supported, while magnetic pressure and tension provide smaller corrections or partial cancellations. As in all the thick accretion disks we study here, the gas pressure supports the thick geometry of the accretion disk.

The MAD force geometry is qualitatively different. The magnitude of the non-geodesic acceleration is larger over much of the inner flow, and the outward radial support acceleration extends into the magnetized surface layers, promoting the development of winds. Both magnetic pressure and magnetic tension contribute substantially to the total acceleration, showing that the inner MAD is not simply a pressure-supported version of a Keplerian disk. Instead, the large-scale magnetic field participates directly in supporting and redirecting the plasma. Magnetic tension becomes dynamically important near the funnel wall and in the inner disk, where curved field lines can provide forces comparable to the gas-pressure term.

The Chimera MAD follows the same broad trend. At early times, its force geometry is intermediate between the SANE and MAD states: the acceleration field is already more magnetically structured than in SANE, but the magnetic terms are not as globally dominant as in the standard MAD. By late times, the magnetic-pressure and tension contributions become more organized and the force geometry moves closer to the MAD state. The late Chimera MAD therefore develops a magnetically regulated support structure while retaining differences in the detailed distribution of acceleration through the disk and surface layers.

The radial profiles in Figure~\ref{fig:force_decomp} make the same point in disk-body averages. As before, positive $a^{\hat r}$ indicates outward radial acceleration and can be interpreted as local support against infall, while negative $a^{\hat r}$ assists the inward plunge. In the SANE flow, the net radial acceleration is modest compared to $R\,\Omega_{\rm Kep}^2$ and is dominated by gas pressure. In the MAD and late Chimera MAD flows, magnetic pressure and tension become comparable to gas pressure, so the inner disk is magnetically regulated rather than simply pressure supported. The magnetic-tension contribution also changes sign across the inner flow, because field-line curvature can either support the gas or assist the plunge depending on location. The interpretation is more complicated near and inside the ISCO, where circular orbital support is no longer the relevant comparison. In the Chimera MAD, the early state has weaker net magnetic support, while the late state develops a radial force budget closer to the standard MAD, though with a distinct balance among gas pressure, magnetic pressure, and magnetic tension.

\subsection{Angular momentum transport}

In an axisymmetric spacetime, azimuthal angular momentum is associated with the Killing direction, and the corresponding conservation law follows from the $\phi$-component of stress-energy conservation
\begin{align}
\partial_t\left(\sqrt{-g}\,T^t{}_\phi\right) + \partial_i\left(\sqrt{-g}\,T^i{}_\phi\right) = 0 ,
\end{align}
where the angular momentum flux ${T^{i}}_\phi$ describes the flux of azimuthal angular momentum through a surface of constant $x^i$ and identifies the channels through which angular momentum is carried. While the poloidal force describes how the gas is supported or accelerated relative to its instantaneous geodesic motion, the divergence of angular momentum flux describes the local torque density, which changes the angular momentum of the flow and allows the infalling matter to change orbits in the first place.

As above, we focus on the poloidal plane and study the $r$ and $\theta$ components of the angular momentum flux. To study the mechanisms that enable transport of angular momentum, we decompose the flux as
\begin{align}
    T^i{}_\phi = \underbrace{w_{\rm fl} u^i u_\phi}_{\rm matter} + \underbrace{b^2 u^i u_\phi}_{\rm EM\ inertia} - \underbrace{b^i b_\phi}_{\rm Maxwell},
    \label{eq:angmom_decomp}
\end{align}
where the first term expresses the angular momentum carried by the fluid inertia in both coherent and incoherent motion, the second term describes the angular momentum carried by moving magnetic energy, and the third is the Maxwell stress arising from magnetic tension when field lines bend. Unlike the scalar shell averages above, this decomposition is applied directly to the angular-momentum flux component $T^i{}_\phi$, so we do not introduce an additional $u^t$ weight.

After averaging over the full azimuthal domain, the $\partial_\phi$ term vanishes identically, and after averaging over a sufficiently long interval in the quasi-steady state the mean time-derivative term is small. In that limit, the remaining poloidal divergence describes the redistribution of the axisymmetrized angular-momentum current through the meridional plane. We therefore interpret the pair $\left(\langle T^r{}_\phi\rangle,\langle T^\theta{}_\phi\rangle\right)$ as a poloidal flux vector of azimuthal angular momentum, and use it to visualize the channels through which the flow changes orbit.

To separate coherent inflow from fluctuating matter transport, we define the enthalpy-Favre average of a quantity $q$ by
\begin{align}
	\langle q\rangle_F \equiv \frac{\langle w_{\rm fl} \, q\rangle}{\langle w_{\rm fl}\rangle},
\end{align}
where $w_{\rm fl}$ is the same fluid enthalpy from before and the angle brackets denote an average over time and azimuth. For our Favre decomposition of the angular momentum fluxes, we restrict to matter-dominated regions by applying a $\sigma<1$ mask so that the decomposition tracks the disk material whose angular momentum must be redistributed for accretion. The matter contribution can then be written as
\begin{align}
	\left\langle w_{\rm fl} u^i u_\phi \right\rangle
	=
	\underbrace{
	\langle w_{\rm fl}\rangle
	\langle u^i\rangle_F
	\langle u_\phi\rangle_F
	}_{\rm mean\ advection} \;\;\;\;\;
	+
	\underbrace{
	{R^i}_\phi
	}_{\rm Favre\ covariance},
	\label{eq:matter_flux_split}
\end{align}
where
\begin{align}
	{R^i}_\phi
	\equiv
	\left\langle w_{\rm fl} u^i u_\phi \right\rangle
	-
	\langle w_{\rm fl}\rangle
	\langle u^i\rangle_F
	\langle u_\phi\rangle_F .
	\label{eq:reynolds_flux_def}
\end{align}
This covariance is the compressible, Favre-weighted analog of a Reynolds contribution. In terms of these pieces, the averaged poloidal angular-momentum flux becomes
\begin{align}
	\left\langle T^i{}_\phi \right\rangle
	&=
	\underbrace{\langle w_{\rm fl}\rangle \langle u^i\rangle_F \langle u_\phi\rangle_F}_{\rm advective} \;\,\;
	+
	\underbrace{ {R^i}_\phi }_{\rm Reynolds} \nonumber  \\[5pt]
	&-
	\underbrace{ \left\langle b^i b_\phi\right\rangle }_{\rm Maxwell} 
	+
	\underbrace{ \left\langle b^2 u^i u_\phi\right\rangle }_{\rm EM\ inertia} .
	\label{eq:full_angmom_flux_split}
\end{align}

The advective term describes the azimuthal angular momentum carried by the mean inflow and mainly reflects the inward movement of the fluid's existing angular momentum rather than the stresses that enable accretion. The Reynolds-like (Favre covariance) term instead measures transport by correlated fluctuations in $u^i$ and $u_\phi$, which can contribute to the redistribution of angular momentum. The Maxwell term measures transport by magnetic tension and is generally the dominant stress-like contribution in magnetized turbulence \citep[e.g.,][]{hawley_1995_localmri,balbus_1998_turbulencetransport,pessah_2006_mristress}. Finally, the electromagnetic-inertia term accounts for angular momentum carried by magnetic energy moving with the flow; this term is typically smaller than the others in dense regions of the disk, but it provides a clean way to account for angular momentum carried by highly magnetized structures.

This decomposition answers a different question from the radial-force analysis. The force decomposition does not directly tell us where angular momentum goes; instead, it describes how the gas is supported or accelerated relative to its instantaneous geodesic motion. The angular-momentum current addresses the complementary question: through which channels is azimuthal angular momentum redistributed? A large Maxwell contribution, for example, indicates efficient magnetic transport of angular momentum, but the corresponding radial magnetic force depends on a different projection and derivative of the same electromagnetic stress-energy tensor.

\begin{figure*}[th!]
\centering
\includegraphics[width=.8\linewidth]{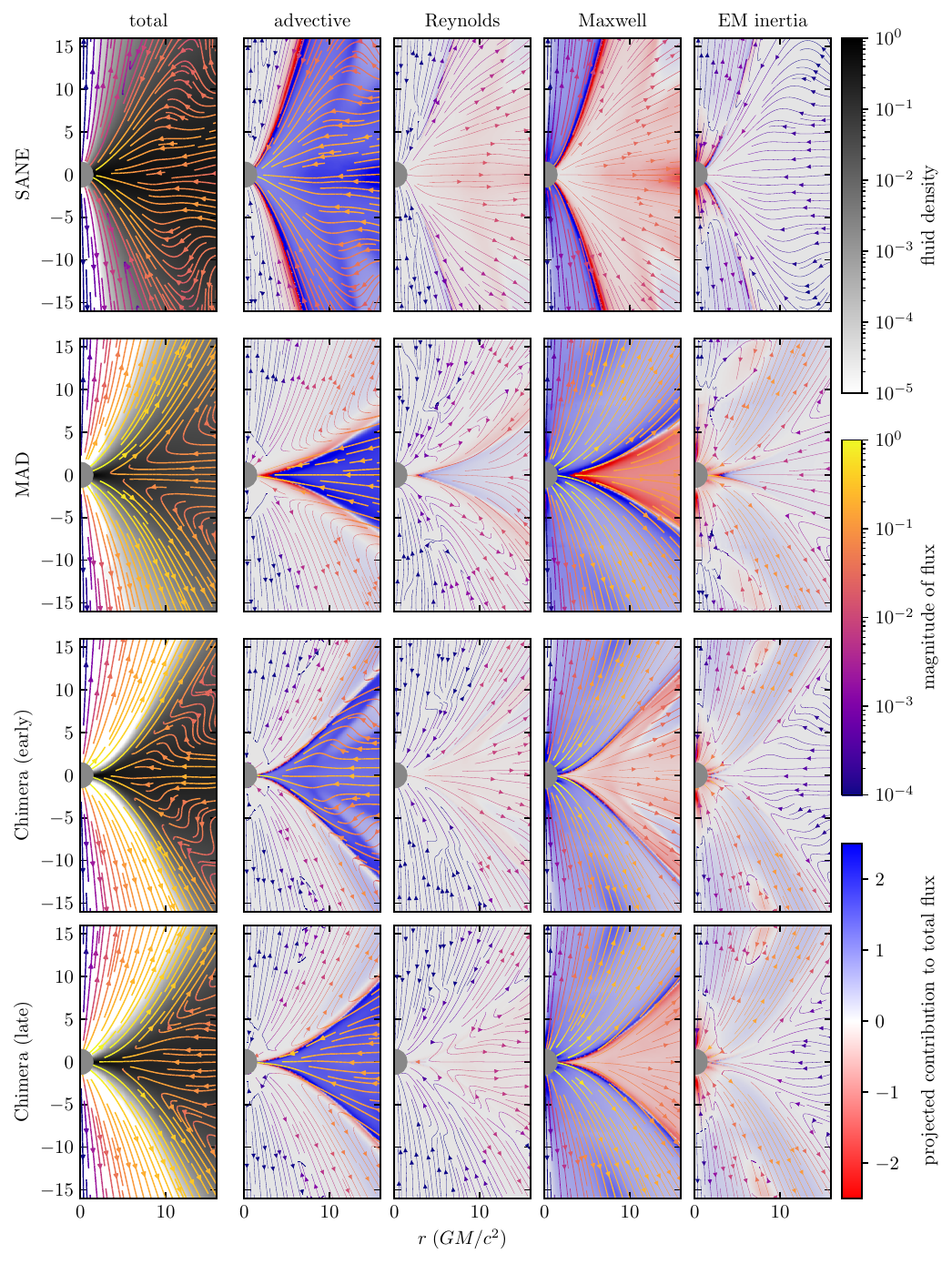}
\caption{
Meridional maps of poloidal angular-momentum flux for the SANE, standard MAD, and Chimera MAD flows, decomposed into total, advective, Reynolds-like (Favre-covariance), Maxwell, and electromagnetic-inertia contributions. Streamlines trace each poloidal flux vector; the color scales show density, current magnitude, and signed projection onto the total-current direction. SANE transport is concentrated in the dense disk body, while MAD and Chimera MAD transport extends into magnetized surface and funnel-wall layers, where Maxwell flux can include both disk/surface-layer torques and electromagnetic angular-momentum extraction from the spinning black hole.
}
\label{fig:angmom_2d}
\end{figure*}

\begin{figure*}[th!]
\centering
\includegraphics[width=\linewidth]{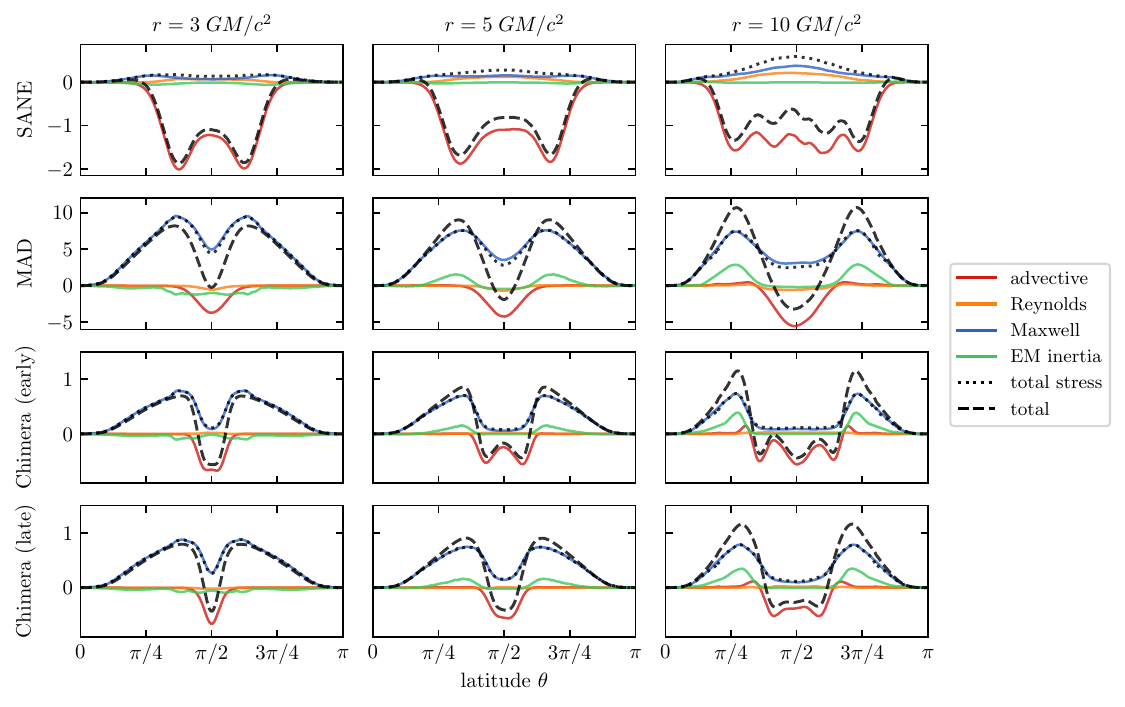}
\caption{
Latitudinal profiles of radial angular-momentum flux at $r = 3$, $5$, and $10\,GM/c^2$, with positive values denoting outward transport. Colored curves show the individual contributions to the total flux (dashed black curve); the dotted black curve shows the sum of the stress-like components ${R^r}_\phi-\langle b^r b_\phi\rangle$. The dotted curve isolates disk transport torques together with possible funnel angular-momentum extraction from the spinning black hole, which together redistribute angular momentum with respect to the fluid. The SANE stress is modest and mostly disk-body confined, the standard MAD is Maxwell dominated with broad off-midplane lobes, and the Chimera MAD follows a similar pattern at reduced amplitude and with different vertical structure.
}
\label{fig:angmom_profiles}
\end{figure*}

Figure~\ref{fig:angmom_2d} shows the geometry of angular momentum transport in the three accretion states. In the SANE flow, the mean current is concentrated in the disk body and is comparatively simple in structure. The dominant transport channels remain close to the midplane, as expected when MRI-driven turbulence redistributes angular momentum primarily through the dense body of the flow. In contrast, in the MAD states, the current becomes more structured in the meridional plane and extends into magnetized surface layers. Once the magnetic field becomes dynamically important, angular momentum need not be transported purely radially through the disk body; it can instead be rerouted poloidally through the corona, surface layers, or eruptive channels. A flow can lose angular momentum and become sub-Keplerian while pressure gradients or magnetic tension provide radial support that prevents immediate plunge. Conversely, a strong radial magnetic force need not imply efficient angular-momentum extraction.

\begin{figure*}[th!]
\centering
\includegraphics[width=\linewidth]{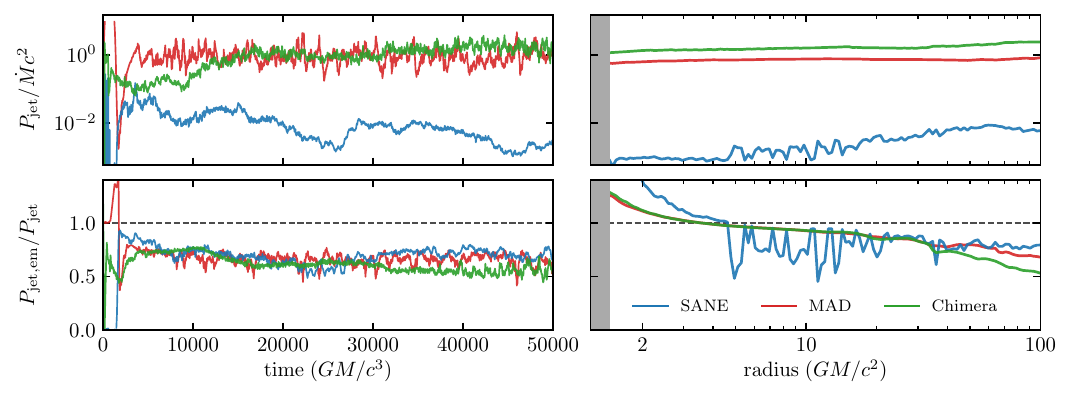}
\caption{
Jet and outflow energetics for the SANE, standard MAD, and Chimera MAD simulations. The jet region is defined from the time- and azimuth-averaged funnel structure, and powers are computed by integrating the rest-mass-subtracted radial energy flux over this region. The left panels show the time evolution of the total jet efficiency and the fractional electromagnetic contribution at $r=100\,GM/c^2$; the right panels show the corresponding radial profiles. The standard MAD and Chimera MAD both produce powerful, predominantly electromagnetic outflows, while the SANE remains much less efficient. The Chimera MAD reaches efficiency $\gtrsim100\%$ by $t\simeq20000\,GM/c^3$, before the clear eruption phase that begins around $t\simeq40000\,GM/c^3$.
}
\label{fig:jet_power}
\end{figure*}

The total angular momentum current identifies the preferred pathways by which angular momentum leaves a given region of the flow, while the decomposition shows which physical channel builds that mean current. In all three models, the magnetic contribution is important, but the balance differs qualitatively. In the SANE, the fluctuating matter contribution is substantial within the disk body and the total transport is comparatively disk-confined. In the MAD states, the Maxwell contribution becomes increasingly prominent, and the transport geometry is more clearly connected to magnetized surface layers. The same distinction appears within the Chimera evolution: Reynolds/Favre-covariance transport contributes in the smoother high-flux interval, while the late eruptive interval is dominated more cleanly by Maxwell transport. The Chimera MAD combines strong magnetic transport with a distinct inner-flow structure in both its smoother high-flux interval and its later eruptive interval.

Figure~\ref{fig:angmom_profiles} shows the latitudinal structure of the radial transport channel at three different radii for a more quantitative visual comparison. The fluid advective contribution is controlled primarily by the vertical profile of the radial velocity (see Figure~\ref{fig:massflux}), with the fluid enthalpy providing a secondary weighting and $u_\phi$ varying more smoothly across the disk body. Because the accreting gas carries its own azimuthal angular momentum inward, we also show the stress-like sum of the Reynolds/Favre and Maxwell terms, ${R^r}_\phi-\langle b^r b_\phi\rangle$, which isolates the stress-like fluxes; in the disk these act as transport torques, while near the funnel they can include electromagnetic angular-momentum extraction from the spinning black hole.

In the SANE flow, the stress-like contribution is small and broad, with Reynolds/Favre and Maxwell transport both confined mostly to the disk body. In the standard MAD, the stress is much larger and almost entirely Maxwell dominated, with broad off-midplane lobes that extend toward the funnel wall. The dashed total current differs from the dotted stress sum where inward advection through the dense disk partially cancels the outward Maxwell stress, especially near the midplane. The Chimera MAD follows the same qualitative pattern as the standard MAD but at reduced amplitude and with a different vertical structure, although the late interval more closely approaches the standard-MAD stress geometry.

\subsection{Jet and outflow energetics}
\label{sec:jet_outflows}

We now study the properties of outflows powered by the black hole accretion flow and spin. Relativistic jets from spinning black holes are commonly understood through the Blandford--Znajek mechanism, in which horizon-threading magnetic flux and frame dragging enable an outward electromagnetic energy flux at the expense of black hole rotational energy \citep{blandford_1977_bz}. This mechanism is one of the standard general relativistic channels for powering observed black hole jets and is largely presumed to be responsible for their energetics \citep[e.g.,][]{eht_m87_5}.

We characterize the jet energetics by first defining the rest-mass-subtracted radial energy flux through a geometrically defined polar funnel. We form the azimuthally averaged radial energy-flux density
\begin{align}
\overline{\mathcal{F}}_{\rm tot}(t,r,\theta) = \left\langle \left(-T^r{}_t-\rho u^r\right)\sqrt{-g} \right\rangle_\phi,
\end{align}
and then integrate this averaged flux over the polar region
$\mathcal{J}$,
\begin{align}
P_{\rm jet}(t,r) = 2\pi\int_{\mathcal{J}} \overline{\mathcal{F}}_{\rm tot}(t,r,\theta)\,d\theta ,
\end{align}
where $\mathcal{J}$ is defined as the region where the average value of $\left\langle \, (\beta\gamma)_\infty\right\rangle_{t,\phi}>1$. We define the electromagnetic contribution $P_{\rm jet}^{\rm EM}$ by replacing $-T^r{}_t-\rho u^r$ in $\overline{\mathcal{F}}_{\rm tot}$ with the electromagnetic component of $-{T^r}_t$ and integrating over the same polar region $\mathcal{J}$. This choice is intended to measure the mean radial energy flux associated with the jet/funnel, including electromagnetic energy transport near the black hole, where the matter velocity can remain inward, so we do not impose a $v^r > 0$ cut. When quoting jet efficiency, we normalize by the instantaneous horizon accretion rate.

We use $(\beta\gamma)_\infty$ as an energy-based terminal-speed proxy to identify the relativistic jet/funnel region since it corresponds to the radial four-velocity at infinity for a steady flow if all of its energy were converted into kinetic energy:
\begin{align}
\gamma_\infty(r,\theta) &\equiv -\frac{\left\langle T^r{}_t \right\rangle_{t,\phi}} {\left\langle \rho u^r \right\rangle_{t,\phi}}, \\
(\beta\gamma)_\infty^2 &\equiv \gamma_\infty^2-1 .
\end{align}

Figure~\ref{fig:jet_power} shows the time evolution and radial structure of these quantities. The SANE model produces a comparatively weak outflow. The standard MAD rapidly reaches high jet efficiency, with substantial variability associated with the same magnetic-flux fluctuations seen in Figure~\ref{fig:horizon_fluxes}. The outflow is predominantly electromagnetic over the inner radial range, indicating that the jet is powered primarily by large-scale magnetic stresses rather than by matter enthalpy or hydrodynamic wind launching.

The Chimera MAD also produces a powerful and predominantly electromagnetic outflow, reaching jet powers comparable to the standard MAD during intervals in which the disk morphology and horizon-flux variability are not those of a saturated, erupting MAD. In particular, the Chimera jet efficiency has already reached values $\gtrsim100\%$, the regime usually associated with MAD spin-energy extraction \citep{tchekhovskoy_2011_mad,narayan_2022_jetsurvey}, by $t\simeq20000\,GM/c^3$, roughly $20000\,GM/c^3$ before the onset of clear eruption activity. The radial profiles reinforce this point. The Chimera MAD outflow is electromagnetically dominated over much of the same radial range as the standard MAD, and its jet efficiency remains far above the SANE value. Thus the Chimera MAD is MAD-like in its ability to launch a strong electromagnetic jet, while its route to strong horizon flux produces a state with a distinct inner-flow and outflow structure. Jet power identifies one important MAD-like diagnostic, but not a unique standard-MAD disk morphology, force balance, or level of eruption activity.

\section{Discussion}
\label{sec:discussion}

We have compared three GRMHD simulations of black hole accretion: a weakly magnetized SANE flow, a standard MAD that rapidly reaches large horizon-threading flux and begins to exhibit clear flux eruptions, and a Chimera MAD. The Chimera simulation accumulates large horizon-threading magnetic flux and launches a powerful electromagnetic jet during an extended high-flux interval before clear eruption activity begins. As more magnetic flux reaches the horizon, the Chimera kinematics move toward the standard-MAD dynamical regime, but its density, temperature, funnel-wall geometry, force balance, angular-momentum transport, and dynamical structure do not collapse onto the standard MAD baseline. The pre-eruptive Chimera interval is therefore not simply a standard MAD caught before its first eruption, and the later eruptive interval is not simply a standard MAD produced by a different transient. Rather, the Chimera represents a long-lived high-flux, high-jet-power accretion history in which several MAD-like diagnostics are present while other dynamical signatures remain distinct.

The comparison also clarifies what the usual SANE and MAD labels do and do not specify. The SANE flow is weakly magnetized by horizon-flux diagnostics and its meridionally resolved mass flux is carried through broad off-midplane channels. The standard MAD rapidly accumulates coherent magnetic flux, evacuates a broader funnel, launches a strong electromagnetic jet, and develops a magnetically reorganized force and transport structure. The Chimera MAD shares parts of both behaviors without matching either limiting case. In this sense, ``magnetically arrested'' is best read as magnetic regulation of the accretion dynamics, including magnetic support and magnetic torques. Horizon magnetic flux is an instantaneous integrated diagnostic; eruption activity is a time-dependent, spatially structured process in which magnetic flux displaces, redirects, and reorganizes the accreting plasma. Jet efficiency, disk morphology, accretion-channel geometry, force balance, meridional transport, and eruption activity should therefore be compared as related but distinct diagnostics rather than treated as consequences of a single scalar state variable. The SANE/MAD distinction remains useful, but accretion history and magnetic-flux supply can shape long-lived accretion phenomenology in ways that matter for how simulations are classified and compared.

The force balance and angular-momentum decompositions show why this separation is physically plausible. Pressure and magnetic-force terms determine how the disk is supported or accelerated, while the angular-momentum current identifies the channels through which gas changes orbit and accretes. A flow can become sub-Keplerian because magnetic stresses remove angular momentum, while magnetic pressure or tension can simultaneously provide radial support or redirect the plasma. Conversely, a strong radial magnetic force does not by itself identify where angular momentum is transported. For the radial angular-momentum flux through the disk, both Maxwell and Reynolds/Favre-covariance terms contribute in the SANE and in the early Chimera interval, whereas the Reynolds term is subdominant in the standard MAD and late Chimera states, where Maxwell transport dominates. In the standard and Chimera MAD-like states, angular-momentum transport extends into magnetized surface layers, funnel-wall regions, winds, and eruptive channels rather than remaining confined to a thin midplane channel \citep[see also][]{chatterjee_2022_fluxangularmomentum,marshall_2018_thinmad,scepi_2024_thinmad,manikantan_2024_madtorques}. The SANE flow can remain weakly magnetized in the horizon-flux sense while accreting through vertically displaced channels, and the Chimera MAD can develop strong magnetic contributions to support and transport without reproducing the standard MAD geometry.

Previous work supports the idea that the MAD label should be read through a force and transport budget, not only through $\phi_{\rm BH}$ or $\dot M$. The original MAD/magnetically choked accretion flow picture emphasized a magnetospheric balance in which accumulated poloidal flux impedes inflow and accretion proceeds through interchange structure \citep{narayan_2003_mad,tchekhovskoy_2011_mad,mckinney_2012_mad}. More recent force-budget analyses show that this balance is geometrically more subtle. In long-duration hot MADs, the flow can be strongly sub-Keplerian even though radial support, especially at intermediate radii, is supplied mainly by thermal pressure and centrifugal terms; poloidal magnetic pressure and tension need not be the dominant outward radial forces, and magnetic tension can partly cancel the outward magnetic-pressure gradient \citep{begelman_2022_whatmad}.

Cooled/thin MAD simulations are consistent: the radial magnetic field does not generally provide radial support to the disk except perhaps very close to the hole, while the vertical structure can become magnetically supported by turbulent magnetic pressure, with wind-driven stresses accelerating inflow and lowering the density \citep{scepi_2024_thinmad}. The corresponding SANE baselines are useful precisely because they remain closer to Keplerian, thermally supported, and radially transported by MRI-like turbulence \citep{narayan_2012_sane,begelman_2022_whatmad,chatterjee_2022_fluxangularmomentum}.

Angular-momentum transport in MADs is likewise not confined to a single midplane channel: thin-MAD studies connect turbulent magnetic stress to eruption activity and find enhanced vertical stresses in coronal or eruptive regions \citep{avara_2016_thinmad,marshall_2018_thinmad,scepi_2024_thinmad}, while long-duration hot MADs show that flux-eruption winds and disk turbulence can both remove angular momentum through surface and wind channels \citep{chatterjee_2022_fluxangularmomentum,manikantan_2024_madtorques}. Similar vertical stratification appears in net-flux and magnetically elevated disk calculations, where magnetic pressure inflates or supports the disk and can shift accretion into surface or coronal layers \citep{miller_2000_magnetizedcorona,zhu_2018_coronalaccretion,mishra_2020_stronglymagnetized,jiang_2019_subeddington}.

This geometry is difficult to compress into the standard $\alpha$-disk picture, where transport is represented by a local effective stress. In the simulations, mass, angular momentum, and magnetic flux move through vertically structured channels. Gas pressure, magnetic pressure, magnetic tension, correlated matter transport, mean advection, and electromagnetic inertia are not generally co-spatial in the meridional plane. This resembles broader work on stratified MRI disks and magnetized accretion flows, where the dense disk body can separate from magnetized surface or coronal layers and transport need not occur in a single midplane region \citep[e.g.,][]{stone_1996_stratified,miller_2000_magnetizedcorona,zhu_2018_coronalaccretion,mishra_2020_stronglymagnetized}. Here the point is not to identify a universal transport channel, but rather to show that vertical stratification allows several routes for mass and angular momentum, with the selected route depending on disk thickness, field geometry, magnetization, funnel evacuation, and the supply of angular momentum and magnetic flux.

The jet results are consistent: a strong electromagnetic outflow identifies coherent large-scale magnetic flux and spin-energy extraction, but it does not uniquely specify the detailed state of the disk that supplied that flux. The Chimera MAD is jet-capable during the extended high-flux interval when its disk morphology and horizon-flux variability differ substantially from the rapidly erupting standard MAD, and it remains distinct after clear eruption activity begins. Jet power is therefore an important MAD-like diagnostic, not a complete classification of the disk dynamics.

\vspace{1em}

We have shown that the route by which magnetic flux is supplied to the inner accretion flow matters. A gas reservoir does not provide anonymous fuel: its size, angular-momentum distribution, and magnetic geometry help determine what reaches the black hole and when. The same point appears in the finite-torus mass-transport picture of \citet{wong_2025_mixing} and in wind-fed \sgra simulations \citep{ressler_2018_stellarwinds,ressler_2020_madstellarwinds}. The Chimera MAD shows that an accretion flow can retain a memory of its feeding mechanism over intervals that would otherwise look quasi-steady in some local diagnostics. That memory should be part of the model interpretation when simulations built from different feeding prescriptions or boundary conditions are compared with one another.

The Chimera MAD is not the only route away from the standard FM-loop SANE/MAD sequence. Thin and radiative MAD studies show that MAD-like magnetic-flux accumulation can coexist with disk structures very different from the hot, thick tori used in many GRMHD model libraries: thin MADs can exhibit enhanced efficiencies, magnetically eruptive activity, magnetically driven winds, vertical magnetic support, coronal dissipation, and wind-driven accretion \citep[e.g.,][]{avara_2016_thinmad,marshall_2018_thinmad,teixeira_2018_radthinmad,scepi_2024_thinmad}. Source-fed models, including stellar-wind-fed \sgra simulations, can develop horizon-scale MAD-like flux accumulation from coherent flux supplied at large radii, while long-duration MAD simulations and intermediate-state/INSANE models emphasize flux eruptions, angular-momentum transport, and transitions between SANE-like and MAD-like behavior \citep[e.g.,][]{ressler_2020_madstellarwinds,chatterjee_2022_fluxangularmomentum,manikantan_2024_madtorques,raha_2026_insane}. Related moving-black-hole/Bondi--Hoyle calculations, potentially relevant to EMRI-like or ambient-flow feeding geometries, also show that large horizon flux and intermittent jet launching can arise without a stationary FM torus \citep[e.g.,][]{kim_2025_bhlmad}. The common lesson is that magnetic state, disk structure, feeding history, transport geometry, and variability occupy a broader space than the original SANE/MAD language by itself suggests.

\vspace{1em}

Our analysis is subject to two important qualifications. First, the standard MAD and Chimera MAD differ in both torus structure and magnetic-field geometry. This is physically motivated by the question of flux supply, but it does not isolate a single control parameter and should be interpreted as a comparison of accretion histories rather than as a controlled parameter sweep. Second, we follow the Chimera simulation for only about $10000\,GM/c^3$ after the onset of the eruptive phase. Evolving the simulation longer could reveal further relaxation, stronger eruption activity, or gradual evolution toward a different late-time state.

\appendix

\section{Force-decomposition details}
\label{sec:force_identity_appendix}

The projected magnetic-stress term in the force equation can be expanded as
\begin{align}
h_{\alpha\nu}\nabla_\mu\left(b^\mu b^\nu\right)
=
h_{\alpha\nu}b^\mu\nabla_\mu b^\nu
+
b_\alpha\nabla_\mu b^\mu .
\end{align}
The induction equation is $\nabla_\mu(u^\mu b^\nu-b^\mu u^\nu)=0$, and contracting this expression with $u_\nu$ and using $u_\nu b^\nu=0$ gives
\begin{align}
\nabla_\mu b^\mu = b^\mu a_\mu ,
\end{align}
so that the second term in the projected magnetic stress is $b_\alpha b_\nu a^\nu$. Moving this term to the left-hand side of the momentum equation gives
\begin{align}
&\left[\left(w_{\rm fl} +b^2\right)h_{\alpha\nu}-b_\alpha b_\nu\right]a^\nu = \\
& \qquad -h_\alpha{}^\nu\nabla_\nu P_{\rm gas} -h_\alpha{}^\nu\nabla_\nu\left(\frac{b^2}{2}\right)
+h_{\alpha\nu}b^\mu\nabla_\mu b^\nu .
\end{align}
In this form, which we use in Equation~\eqref{eq:force_balance_general}, the divergence term is thus absorbed into the anisotropic inertia tensor multiplying the acceleration.

\section{Orthonormal projections}
\label{sec:coordinate_basis}

We evaluate the force decomposition in Kerr--Schild coordinates, but the coordinate basis directions $\partial_r$ and $\partial_\theta$ are not guaranteed to be spatial directions in the fluid rest frame. We therefore construct the $\hat{r}$ and $\hat{\theta}$ directions used in Figures~\ref{fig:acceleration_2d_all}--\ref{fig:force_decomp} by projecting the coordinate basis vectors into the space orthogonal to the local fluid four-velocity and then orthonormalizing them.

The (mixed) projection tensor into the fluid frame is
\begin{align}
    {h^\mu}_\nu = \delta^\mu{}_\nu + u^\mu u_\nu .
\end{align}
For a coordinate direction labeled $X = r, \theta$, the rest-frame projection of the coordinate basis vector ${\delta^\mu}_X$ is
\begin{align}
    \tilde e^\mu_X &= h^\mu{}_\nu \delta^\nu{}_X \\
    &= {\delta^\mu}_X + u^\mu u_X ,
\end{align}
with norm (no sum over the label $X$)
\begin{align}
    g_{\mu\nu}\tilde e^\mu_X \tilde e^\nu_X = g_{XX}+u_X u_X .
\end{align}
The radial basis vector is therefore
\begin{align}
    e^\mu_{\hat r} = \dfrac{\delta^\mu{}_r + u^\mu u_r}{\left(g_{rr}+u_r u_r\right)^{1/2}} ,
    \label{eq:e_hatr}
\end{align}
and the analogous $\theta$-directed vector is
\begin{align}
    e^{\mu}_{\hat\theta,0} = \dfrac{\delta^\mu{}_\theta + u^\mu u_\theta}{\left(g_{\theta\theta}+u_\theta u_\theta\right)^{1/2}} .
\end{align}

Although $e^{\mu}_{\hat\theta,0}$ is spatial in the fluid frame and has unit norm, it is not generally orthogonal to $e^\mu_{\hat r}$. We therefore remove its component along $e^\mu_{\hat r}$,
\begin{align}
    \bar e^\mu_{\hat\theta} = e^{\mu}_{\hat\theta,0} -  e^\mu_{\hat r} \; g_{\alpha\beta} \, e^\alpha_{\hat r} \, e^{\beta}_{\hat\theta,0} ,
\end{align}
and renormalize,
\begin{align}
    e^\mu_{\hat\theta} = \frac{\bar e^\mu_{\hat\theta}} {\left(g_{\alpha\beta} \bar e^\alpha_{\hat\theta} \bar e^\beta_{\hat\theta}\right)^{1/2}} .
    \label{eq:e_hattheta}
\end{align}
The two meridional basis vectors then obey
\begin{align}
    u_\mu e^\mu_{\hat A} &= 0 ,
    \\
    g_{\mu\nu} e^\mu_{\hat A} e^\nu_{\hat B}
    &= \delta_{\hat A\hat B},
    \qquad
    \hat A,\hat B \in \{\hat r,\hat\theta\}.
\end{align}

For any covariant acceleration $a_\mu$, the physical meridional components used
in the plots are the contractions with these orthonormal basis vectors:
\begin{align}
    a^{\hat r} &= e^\mu_{\hat r} a_\mu , \\
    a^{\hat\theta} &= e^\mu_{\hat\theta} a_\mu .
    \label{eq:ahat_def}
\end{align}
The same projection is applied to each term in the force decomposition as needed:
\begin{align}
    a^{\hat A}_{\rm gas} &= e^\mu_{\hat A} a^{\rm gas}_\mu , \\
    a^{\hat A}_{\rm magP} &= e^\mu_{\hat A} a^{\rm magP}_\mu , \\
    a^{\hat A}_{\rm tens} &= e^\mu_{\hat A} a^{\rm tens}_\mu , \qquad \hat A\in\{\hat r,\hat\theta\}.
    \label{eq:ahat_components}
\end{align}

\bibliography{main}
\bibliographystyle{aasjournal}

\end{document}